\documentclass[apj, singlecolumn]{aastex63}
\usepackage{epstopdf}
\usepackage{graphicx}
\usepackage{wrapfig}
\usepackage[utf8]{inputenc}
\usepackage{amsmath}
\usepackage{natbib}
\usepackage{multirow}
\usepackage{longtable}
\bibpunct{(}{)}{;}{a}{}{,}
\usepackage{comment}
\usepackage{longtable}

\usepackage{color}
\definecolor{darkgreen}{RGB}{0,142,128}
\definecolor{darkblue}{RGB}{0,100,170}
\definecolor{darkpurple}{RGB}{150,0,150}

\usepackage{hyperref}
\hypersetup{colorlinks,citecolor=blue,linkcolor=blue}

\begin{document}
\title{Modeling Solar Wind Variations over an 11-yr Cycle with Alfvén Wave Dissipation: a Parameter Study}
\author[0000-0002-7069-1711]{Soumitra Hazra}
\affil{Département d'Astrophysique/AIM, CEA/IRFU, CNRS/INSU, Université de Paris-Saclay, Université de Paris, CEA Paris-Saclay, 91191 Gif-sur-Yvette, France}
\affil{Université Paris-Saclay, CNRS,  Institut d'astrophysique spatiale, 91405, Orsay, France}
\email{soumitra.hazra@cea.fr, soumitra.hazra@gmail.com}
\author[0000-0002-2916-3837]{Victor Réville}
\affiliation{IRAP, Université Toulouse III - Paul Sabatier,
CNRS, CNES, Toulouse, France}
\email{victor.reville@irap.omp.eu}
\author[0000-0002-2137-2896]{Barbara Perri}
\affil{Université Paris-Saclay, CNRS,  Institut d'astrophysique spatiale, 91405, Orsay, France}
\affil{Département d'Astrophysique/AIM, CEA/IRFU, CNRS/INSU, Université de Paris-Saclay, Université de Paris, CEA Paris-Saclay, 91191 Gif-sur-Yvette, France}
\email{barbara.perri@cea.fr}
\author[0000-0002-9630-6463]{Antoine Strugarek}
\affil{Département d'Astrophysique/AIM, CEA/IRFU, CNRS/INSU, Université de Paris-Saclay, Université de Paris, CEA Paris-Saclay, 91191 Gif-sur-Yvette, France}
\email{antoine.strugarek@cea.fr}
\author[0000-0002-1729-8267]{Allan Sacha Brun}
\affil{Département d'Astrophysique/AIM, CEA/IRFU, CNRS/INSU, Université de Paris-Saclay, Université de Paris, CEA Paris-Saclay, 91191 Gif-sur-Yvette, France}
\email{sacha.brun@cea.fr}
\author[0000-0003-4290-1897]{Eric Buchlin}
\affil{Université Paris-Saclay, CNRS,  Institut d'astrophysique spatiale, 91405, Orsay, France}
\email{eric.buchlin@ias.u-psud.fr}
\begin{abstract}
We study the behaviour and properties of the solar wind using a 2.5D Alfvén wave driven wind model. We first systematically compare the results of an Alfvén wave (AW) driven wind model with a polytropic approach. Polytropic magnetohydrodynamic wind models are thermally driven, while Alfvén waves act as additional acceleration and heating mechanisms in the Alfvén wave driven model. We confirm that an AW-driven model is required to reproduce the observed bimodality of slow and fast solar winds. We are also able to reproduce the observed anti-correlation between the terminal wind velocity and the coronal source temperature with the AW-driven wind model. We also show that the wind properties along an eleven year cycle differ significantly from one model to the other. The AW-driven model again shows the best agreement with observational data. Indeed, solar surface magnetic field topology plays an important role in the Alfvén wave driven wind model, as it enters directly into the input energy sources via the Poynting flux. On the other hand, the polytropic wind model is driven by an assumed pressure gradient; thus it is relatively less sensitive to the surface magnetic field topology. Finally, we note that the net torque spinning down the Sun exhibits the same trends in the two models, showing that the polytropic approach still captures correctly the essence of stellar winds. 
\end{abstract}

\section{Introduction}
It is now assumed that the large-scale magnetic field of most solar-like stars is created by large scale convective flows acting inside the stellar interior \citep{brun15, shaz16}. The solar large-scale magnetic field varies cyclically with an average periodicity of eleven years; it is known as the solar cycle. When the Sun is at the maximum of the solar cycle, there are large number of solar storms (flares and coronal mass ejections) which affect the space and ground based vulnerable infrastructures. Solar magnetic field is multipolar and complex at the cycle maximum, while it is dipolar at the cycle minimum. This large scale magnetic field is responsible for the structure of the solar or stellar corona. \cite{park58} argued that it is not possible to have a static equilibrium corona due to the large temperature of the solar corona and low pressure of the interstellar medium. A supersonic expansion driven by the pressure gradient (known as solar wind) is then the only possible steady-state for the outer solar atmosphere. Later, standard magnetohydrodynamic (MHD) solar and stellar wind models have been developed considering the effect of both rotation and the magnetic field, also known as the theory of magnetic rotators \citep{schatz62, park63, webe67, mest68}. As the solar wind is accelerated outward from the solar surface, it carries information regarding the complex topology of the solar magnetic field \citep[see for instance ][]{wang03, pint11, revi15}. The solar wind indeed exhibits significant variation in speed and density and has a spiral shape due to 28-days solar rotation. There is also a large variation in the density and speed of the interplanetary medium during solar storms. The solar wind is important from the Sun's evolution point of view as it carries angular momentum away from the Sun \citep{webe67, schatz59, schatz62}. Because of the magnetized nature of the solar wind, it also exerts a magnetic torque that slows down the Sun \citep{skum72}. In summary, it is necessary to have a clear understanding of the behaviour and structure of the solar wind for the perspective of space weather and stellar evolution. 

Various in situ spacecraft also indicate that the structure and speed of the solar wind are largely controlled by the solar magnetic activity. Ulysses' first orbit observations showed that near the solar minimum, the solar wind has a relatively simple bimodal structure in the outer heliosphere. Fast solar wind emanates from the high latitude polar coronal holes, while slower, high-density solar wind is found at low latitudes \citep{mcco98a, mcco20a}. Ulysses' second orbit, which occurred over the maximum of cycle 23, found a complex magnetic structure of the solar wind at all latitudes \citep{mcco02, mcco02a}. The complex solar wind structure at the solar maximum is primarily due to the mixture of different flows arising from different sources like streamers, coronal holes, active regions, etc, at various latitudes \citep{mcco02, mcco08, neug02}. However, the reason behind the behaviour of the solar wind plasma in the solar corona is still unclear. 


Different studies have been performed to explain Ulysses' observations by solving MHD equations in two and three dimensions with the observed photospheric vector magnetogram as a boundary condition \citep{kepp99, usma20, usma03, pint11, vand10, toth12, rile15,revi17}. Many of these studies assumed that the solar wind is driven by the pressure gradient from an approximately $10^6$K corona, which is modeled through a polytropic equation of state \citep{wash93, kepp99, matt12, revi15, revi15a}. These polytropic MHD models are also successful in explaining some of the observational properties of the solar wind \citep{pint11, matt15, revi16, revi17, perr18, finl18, finl18a}. However, it was shown that coronal density has to be sufficiently low to match the observed interplanetary density \citep{hund72, leer82, stew97}. If one assumes the observed coronal density, a polytropic wind will lead to an interplanetary density too high compared to observations \citep[see][]{hund72}. In summary, we need an additional source of energy and momentum to produce a relatively low-density fast wind while keeping the coronal density consistent with observations \citep{munr77,barn95}. \cite{rile01} used an ad-hoc mechanism to incorporate an additional source of acceleration in their solar wind model and found good agreement with Ulysses' observations. 

In recent times, Alfvén waves have been proposed to be one of the best mechanisms for the additional acceleration of the solar wind plasma \citep{belc71, jacq77}. While the Alfvén wave pressure acts as a mechanism for the solar wind acceleration, the dissipation of waves can be responsible for heating \citep{holl86}. The Helios 1 and 2 missions have given us detailed observations regarding the behaviour of the velocity and magnetic field fluctuations ($\delta v$ and $\delta B$ respectively) in the solar system. Peculiar radial evolution of the $\delta v$ and $\delta b$ at the time of maximum cross helicity can be explained by the properties of outward propagating Alfv\'en waves \citep{tu93, tu95}. Cross helicity basically describes the cross-correlation between the fluid velocity fluctuations ($\delta v$) and magnetic field fluctuations ($\delta b$) \citep{yoko11}. Observations from these missions indicate that fluctuations of the velocity and magnetic field have clear radial evolution and they are highly correlated \citep{mars90, tu93}. Both their correlation and their radial evolution can be explained by the properties of outward propagating Alfvén waves \citep{belc71a}. High-resolution observations also found the signature of Alfvén waves in the solar atmosphere \citep{depo07, tomc07}. 

One can characterize the radial evolution of magnetic and velocity field fluctuations by different parameters, namely the fluctuation energy, cross helicity, and the Alfvén ratio \citep{matth82}. Alfvén ratio basically represents the ratio between the kinetic and magnetic fluctuation energy. Wentzel-Kramers-Brillouin (WKB) theory explains the decrease of fluctuation energy with distance as a consequence of the Alfvén wave propagation in the solar wind \citep{holl74, mars90, tu93, tu95}. It is thus necessary to include the basic essence of the Alfvén wave propagation and dissipation mechanisms in the solar wind model. In the Alfvén wave driven solar wind model, footpoints of flux tubes are shaken randomly by the convection at the solar surface and generate wave-like fluctuations (usually Alfvén waves) that propagate up to the corona. It is also believed that these waves are partially reflected back toward the Sun and generate MHD turbulence that is responsible for the gradual wave energy dissipation into the corona \citep{vell89, cran12}. Several one\nobreakdash-, two\nobreakdash-, and three-dimensional solar wind models with Alfvén wave propagation and dissipation mechanism have been developed over the years to explain different properties of the solar wind \citep{tu93, usma20, verd09, verd10, soko13, vand14, lion14, revi18, shod18, shod20}.

In this paper, we aim to characterize and test the behaviour and properties of the Alfvén wave driven solar wind model \citep{revi20a} over an eleven-year solar cycle. So far this particular model has been tested for only one epoch and only on the Parker Solar Probe (PSP) trajectory, showing an encouragingly excellent comparison with in-situ observations. Please note that we denote the MHD solar wind model with Alfvén wave propagation and dissipation mechanisms as {\it Alfvén wave driven solar wind model} (AW model) for the rest of our paper. One can see \cite{vand10, vand14, soko13, lion14, cohe17, revi20, revi20a} for similar approaches. Here we model the Alfvén waves propagation and dissipation mechanisms following the WKB theory \citep{alaz71, belc71, whan73, holl74} which consists in solving two extra equations --- one for parallel Alfvén waves ($\delta v$ is in the direction of $\delta B$) and another for antiparallel Alfvén waves ($\delta v$ is in the opposite direction than $\delta B$). We furthermore use the Wilcox synoptic maps of photospheric magnetic field as an inner boundary of our model. In this study, we first compare the properties of the solar wind structures obtained from the polytropic and AW model to see how the AW model performs compared to the polytropic one. Next, we try to understand the impact of various coronal parameters on the AW model solution for few specific time period. Finally, we test the property of the AW model over the cycle 23.

Section~\ref{sec:models} describes the details of our solar wind models, both polytropic and AW. We present a detailed comparison between the results obtained from both models in Section~\ref{sec:comparison}. We also investigate there the relationship between solar wind terminal velocity and the coronal source temperature. In Section~\ref{sec:dependence}, we characterize how the mass loss, angular momentum loss, and averaged Alfvén radius depend on the coronal parameters in the AW model. We also derive a simple scaling law for the mass loss and the net torque applied to the star \citep{revi15a}. In Section~\ref{sec:loss}, we study the variation of mass and angular momentum loss during cycle 23. Finally, we present a summary of this study and our conclusions in the last section.

\section{Solar Wind models}
\label{sec:models}

In this section, we provide the basic governing equations for the two classes of models used in this study, namely the polytropic wind model and the Alfvén wave driven MHD solar wind model. We use the PLUTO MHD code \citep{mign07} to solve the equations described in the following. 

\subsection{Polytropic MHD Solar Wind Model}
\label{sec:modpoly}

In the polytropic model, the MHD equations are written as usual in their conservative form:

\begin{equation}
\label{eq:poly1}
\frac{\partial}{\partial t} \rho + \nabla \cdot \rho \mathbf{v} = 0,
\end{equation}
\begin{equation}
\label{eq:poly2}
\frac{\partial}{\partial t} \mathbf{m} + \nabla \cdot ( \mathbf{mv}-\frac{1}{\mu_0} \mathbf{BB}+\mathbf{I}p) =  - \rho \nabla \Phi ,
\end{equation}
\begin{equation}
\label{eq:poly3}
\frac{\partial}{\partial t} (E + \rho \Phi)   + \nabla \cdot ((E+p+\rho \Phi)\mathbf{v}-\frac{1}{\mu_0} \mathbf{B}(\mathbf{v} \cdot \mathbf{B})) = 0,
\end{equation}
\begin{equation}
\label{eq:poly4}
\frac{\partial}{\partial t} \mathbf{B} + \nabla \cdot (\mathbf{vB}-\mathbf{Bv})=0,
\end{equation}
where $\mathbf{B}$ is the magnetic field, $\rho$ is the mass density, $\mathbf m = \rho \mathbf v$ is the momentum, $E = \rho e + \rho v^2/2 + B^2/2$ is the total energy, $\mathbf{v}$ is the velocity field, $p = p_{\mathrm{th}} + B^2/2$ is the total (thermal and magnetic) pressure and $\mathbf{I}$ is the identity matrix.

So called polytropic models \citep{wash93, kepp99, matt12, revi15} are build using the ideal equation of state $\rho \epsilon = p_{th} / (\gamma -1)$, where $\epsilon$ represents the internal energy per mass and $p_{th}$ is the thermal pressure. The adiabatic exponent $\gamma$ is then chosen to be $<5/3$, which is the value expected for an adiabatic expansion of the solar wind. Decreasing $\gamma$ artificially is equivalent to creating an upward free-streaming heat flux ($\mathbf q \propto p_{th} \mathbf v$), down to the (unreached) limit $\gamma=1$, where the heat flux is infinitely fast, and the plasma isothermal. Polytropic winds are thus taking advantage of a value of $\gamma$ close to one (in the range $[1.05,1.2]$), in combination with an already hot corona at the base of the domain, to create the pressure gradient necessary to the flow acceleration. Our solar wind model uses the value $\gamma=1.05$ and is initialized with the solution of a one-dimensional, hydrodynamic polytropic wind. 


We perform 2.5D axisymmetric (around the rotation axis of the Sun) polytropic wind simulations in the meridional plane $R_\odot \le r \le 20 R_\odot$ and $0 \le \theta \le  \pi$ with a resolution of $512 \times 512$ (i.e. $N_r =N_\theta =512$). Please note that we use uniform grid in latitude and stretched grid in radius. Our grid spacing in radius varies from 0.001 solar radii at the surface of the star to 0.01 solar radii at the outer boundary. We use the Harten, Lax, van Leer Riemann solver \citep[HLL, see][]{einf88} to solve the MHD equations in spherical polar coordinates. We maintain $\nabla \cdot \mathbf{B} = 0$ using an hyperbolic divergence cleaning method \citep{dedn02}. We describe the initial and boundary conditions for both polytropic and Alfvén wave driven solar wind model in Sections~\ref{sec:initial} and~\ref{sec:boundary} respectively. 

\subsection{Alfvén Wave driven MHD Solar Wind Model}
\label{sec:modalfven}

Alfvén wave driven models aim to account for both the coronal heating and the additional source of energy and momentum necessary to create the observed velocity bimodality. The turbulent dissipation, a consequence of the observed cascade \citep[see, e.g.][]{tu93,tu95}, is one of the main factor of the coronal heating in these models, while the Alfvén wave pressure \citep{alaz71,belc71} comes as an additional acceleration term for the wind. 

The model used in this work is fully described in \citet{revi20a}, we will in this section focus on the differences with the polytropic model. First, two equations are solved in addition to the usual MHD equations. These equations read : 

\begin{equation}
    \frac{\partial \mathcal{E}^\pm}{\partial t} + \nabla \cdot \left( [\mathbf{v} \pm \mathbf{v_A}] \, \mathcal{E}^\pm \right) = -\frac{\mathcal{E}^\pm}{2} \nabla \cdot \mathbf{v}  - Q_w^\pm,
\end{equation}
where $\mathbf {v_A}$ is the Alfvén speed and
\begin{equation}
    \mathcal{E}^\pm =  \rho \frac{|z^\pm|^2}{4}
\end{equation}
is the energy density for parallel and antiparallel Alfvénic perturbations in their Elsässer form:

\begin{equation}
    \mathbf z^{\pm} = \delta \mathbf v \mp \mathrm{sign} (\mathbf B_r)\frac{\delta \mathbf b}{\sqrt{\mu_0 \rho}}.
\end{equation}

where $\delta \mathbf b$ and $\delta \mathbf v$ is the fluctuations in the magnetic field and velocity field (transverse velocity) respectively. These equations essentially follow the Wentzel-Kramers-Brillouin theory \citep[WKB][]{alaz71,belc71}, with a dissipation term $Q_w$ is defined as:

\begin{equation}
Q_w = Q_w^+ + Q_w^-,
\end{equation}
where each term due to different populations of Alfvén waves is prescribed as:
\begin{equation}
Q_w^\pm = \frac{\rho}{8}  \frac{|z^\pm|^2}{\lambda} \left(\mathcal{R} |z^\pm| + |z^\mp|\right),
\label{eq:kol_ph2}
\end{equation}
which follows the Kolmogorov phenomenology. It was previously suggested that Kolmogorov type dissipation can take place only if there are counter-propagating Alfvén waves \citep{cran05, chan09}. In coronal holes, partial reflection due to inhomogeneties in the solar corona are responsible for these counter-propagating Alfvén waves \citep{verd09}. Keeping these facts in mind, we set a constant reflection coefficient $\mathcal{R}=0.1$ for the sake of simplicity. This coefficient is close to what is obtained in the analytical model of \cite{chan09a}, which leads to very similar heating rates. $\lambda$ is the dissipation length scale and varies with distance to the Sun as the inverse of the square root magnetic field strength (or the width of a given flux tube). The value at the base of the corona is set to $\lambda_{\odot} = 0.025 R_{\odot}$, which is the size of supergranules \citep[see][for a more detailed discussion]{revi20}.

These equations are coupled to the rest of the MHD system through two terms. First, we add in the total pressure the contribution of Alfvén waves: 

\begin{equation}
    p = p_{\mathrm{th}} + B^2/2+ \mathcal{E}/2,
\end{equation}
where $\mathcal{E} = \mathcal{E}^+ + \mathcal{E}^-$. This terms acts, in particular, directly in the momentum equation to add an additional outward pressure gradient. Then, we modify the energy equation that becomes:

\begin{eqnarray}
\label{eq:alfven3}
\frac{\partial}{\partial t} (E + \mathcal{E} + \rho \Phi)  + \nabla \cdot [(E+p+\rho \Phi)\mathbf{v} \nonumber \\
-\mathbf{B}(\mathbf{v} \cdot \mathbf{B}) +
\mathbf v_g^+\mathcal{E}^+ + \mathbf v_g^-\mathcal{E}^- ] = Q,
\end{eqnarray}
where $\mathbf v_g^{\pm} = \mathbf v \pm \mathbf v_A$ denotes the group velocity of the Alfvén wave packets. Here $\gamma = 5/3$, which calls for a new term $Q$, that does not exist in the polytropic model, to control the thermodynamics of the Alfvén wave driven model. It can be decomposed in four contributions (sources and sinks):
\begin{equation}
\label{eq:q}
Q =  Q_w + Q_h - Q_c - Q_r.
\end{equation}

The sources are made of the Alfvén wave heating term $Q_w$ described above, and an ad-hoc term following \citet{with88}:
\begin{equation}
Q_h = F_h/H \left(\frac{R_{\odot}}{r} \right)^2 \exp{ \left(-\frac{r-R_{\odot}}{H}\right)},
\end{equation}
with $H \sim 1 R_{\odot}$, the heating scale-height, and $F_h$ is the energy flux from the photosphere (in $\mathrm{erg \cdot cm^{-2}s^{-1}}$). This additional term is used to render heating in the chromosphere and lower corona that could find a different origin than Alfvén waves \citep[nanoflares for instance, see][]{park88}. 

The sinks are made of the usual thin radiation cooling term
\begin{equation}
Q_r = n^2 \Lambda (T),
\label{eq:radloss}
\end{equation}
with $n$ the electron density, $T$ the electron temperature, and $\Lambda(T)$ defined as in \citet{atha86}, and of the thermal conduction flux $Q_c$. The latter combines a collisional and a collisionless prescription:
\begin{equation}
Q_c = \nabla \cdot (\alpha \mathbf{q}_s + (1-\alpha) \mathbf{q}_p),
\end{equation}
$\mathbf{q}_s = -\kappa_0 T^{5/2} \nabla T$ being the usual Spitzer-Härm collisional thermal conduction with $\kappa_0=9 \times 10^{-7}\;\mathrm{cgs}$, and $\mathbf{q}_p = 3/2 p_\mathrm{th} \mathbf{v}_e$ the free-stream heat flux \citep{holl86}. The coefficient $\alpha = 1 / (1 + (r - R_\odot)^4 / (r_\mathrm{coll} - R_\odot)^4)$ creates a smooth transition between the two regimes at a characteristic height of $r_\mathrm{coll} = 5 R_\odot$.

As in the polytropic model, we performed all our axisymmetric Alfvén wave driven solar wind simulations in the meridional slab $R_\odot \le r \le 20 R_\odot$ and $0 \le \theta \le \pi$ with a resolution of $256 \times 256$ (i.e. $N_r  = N_\theta = 256$). We use uniform grid in the latitudinal direction and a stretched grid in the radial direction. Note that resolution is different in the AW and polytropic scenario. The equations are solved using an improved Harten, Lax, van Leer Riemann solver \citep[HLL, see][]{einf88}, combined with a parabolic reconstruction method and minmod slope limiter. We  use an hyperbolic divergence cleaning method \citep{dedn02} to maintain $\nabla \cdot \mathbf{B} = 0$. 

\subsection{Initial Conditions}
\label{sec:initial}
 We initialize our solar wind models with the solutions from a hydrodynamic polytropic wind and from a Potential Field Source Surface (PFSS) magnetic field extrapolations \citep{scha69}  obtained from Wilcox solar magnetic maps. 

 \begin{figure}[!ht]
\centering
\begin{tabular}{cc}
\includegraphics*[width=1.0\linewidth]{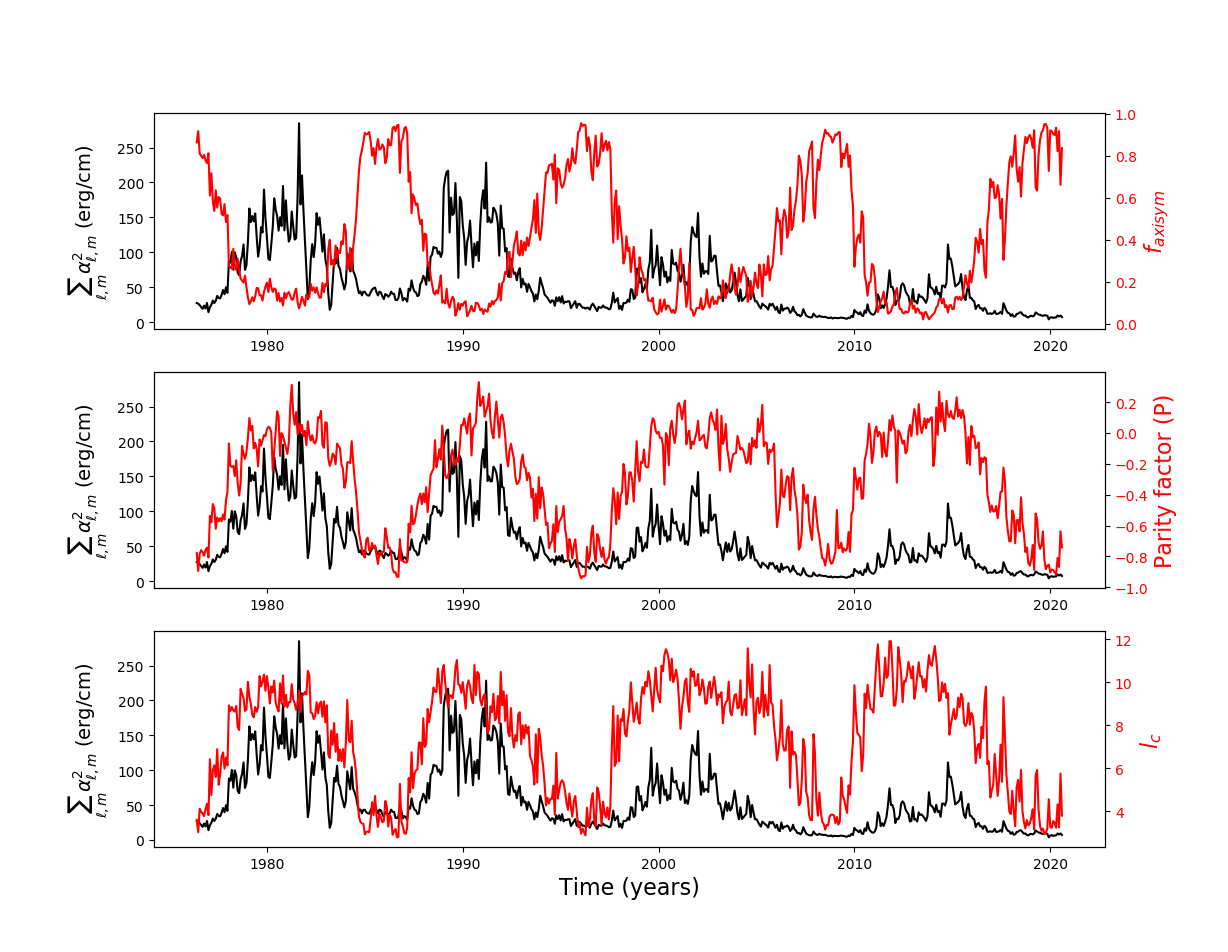}
\end{tabular}                
\caption{\footnotesize{Top panel: Comparison between the total surface magnetic energy $\sum_{\ell,m} \alpha_{\ell,m}^2$ (black line) and the axisymmetric component $f_\textrm{axisym}$ obtained from the Wilcox magnetic maps (red line). Middle panel: Comparison between the total surface energy $\sum_{\ell} \alpha_{\ell,m}^2$ (black line) and the parity factor ($P$) obtained from the Wilcox magnetic maps (red line). Negative parity factor indicates that surface magnetic field is mostly antisymmetric across the equator; while positive indicates opposite. Bottom panel: Comparison between the total surface energy $\sum_{\ell} \alpha_{\ell,m}^2$ (black line) and the map complexity ($l_c$) obtained from the Wilcox magnetic maps (red line). It indicates that complexity is highly correlated with the total surface energy.}}
\label{fig:surfen}
\end{figure}

Wilcox Solar Observatory (WSO) provides continuous line of sight magnetogram observations since May 27, 1976 \citep{sche77}. We decompose the Wilcox surface radial magnetic field at a given time $t_0$ in terms of spherical harmonic coefficients for each Carrington rotation (CR), as in \cite{dero12}:
\begin{equation}
 \label{eq:br}
     B_r(R_\odot, \theta, \phi, t_0)=\sum_{\ell,m} \alpha_{\ell,m} (R_\odot,t_0)  Y_{\ell,m} (\theta, \phi)
 \end{equation}
 
where $Y_{\ell,m}$ are the normalized spherical harmonics. We use $\ell_{\mathrm{max}} = 40$ for the decomposition. Figure~\ref{fig:surfen} shows the properties of the solar surface magnetic energy ($\sum_{\ell,m} \alpha_{\ell,m}^2$) for the last 45 years. We define a parameter, named $f_\textrm{axisym}$, to quantify the energy of the axisymmetric component over the total surface energy following the equation:
 \begin{equation}
     f_\textrm{axisym}= \frac{\sum_{\ell} \alpha_{\ell,0}^2}{\sum_{\ell,m} \alpha_{\ell,m}^2},
 \end{equation}
 
 Similarly, we also define a quantity
 \begin{equation}
     f_\textrm{non-axisym}= \frac{\sum_{\ell,m\ne 0} \alpha_{\ell,m}^2}{\sum_{\ell,m} \alpha_{\ell,m}^2} = 1 - f_\textrm{axisym},
 \end{equation}
 to quantify the energy of the non-axisymmetric component over the total surface energy ($\frac{1}{2\mu_0}\sum_{\ell,m} \alpha_{\ell,m}^2$). Figure~\ref{fig:surfen} shows the variation in time of the total surface magnetic energy. We observe a clear anti-correlation between the total surface magnetic energy and $f_\textrm{axisym}$ over the whole period.  It simply means axisymmetric surface energy is the main contributor to the total surface energy (typically more than 80\%) at the cycle minimum. On the other hand, a major contribution to the total energy (typically more than $80 \%$) comes from the non-axisymmetric components of the solar magnetic fields at the solar maximum (as $f_\textrm{non-axisym}= 1- f_\textrm{axisym}$). In summary, non-axisymmetric surface energy is the major part of the total surface energy at the cycle maximum; while axisymmetric surface energy is at the cycle minimum.
 
 We have also defined two other quantities, namely, the parity factor ($P$) and the map complexity to characterize the properties of the synoptic magnetic maps:
  \begin{equation}
     P = \frac{\sum_{\ell+m=even} \alpha_{\ell,m}^2- \sum_{\ell+m=odd} \alpha_{\ell,m}^2}{\sum_{\ell,m} \alpha_{\ell,m}^2} ,
 \end{equation}
where $\sum_{\ell+m=even} \alpha_{\ell,m}^2$ corresponds to the symmetric surface energy with respect to the equator and $\sum_{\ell+m=odd} \alpha_{\ell,m}^2$ corresponds to the antisymmetric surface energy with respect to the equator. When parity factor ($P$) is closer to 1, the magnetic field is more symmetric; closer to -1, it is more antisymmetric. Parity actually quantifies the non-linear coupling between symmetric and antisymmetric modes of solar magnetic fields across the hemisphere \citep[see][for more details]{dero12,shaz19}. Please note that if one considers axisymmetry and ignores higher order modes, then parity basically represents the non-linear coupling between the dipolar and quadrupolar modes of solar magnetic fields across the hemisphere. The middle panel of Figure ~\ref{fig:surfen} shows the evolution of the parity function with time. We notice that the parity factor is mostly negative throughout the time; that indicates the antisymmetric nature of the magnetic field across the equator \citep{dero12}. 
We define the characteristic harmonic degree as:
\begin{equation}
     l_c = \frac{\sum_{\ell,m} \ell ~\alpha_{\ell,m}^2}{\sum_{\ell,m} \alpha_{\ell,m}^2} .
 \end{equation}
 A higher $l_c$ generally indicates the presence of more complex magnetic fields at the solar surface. The bottom panel of Figure \ref{fig:surfen} shows the time evolution of complexity. If we compare the time evolution of complexity with the time evolution of total surface energy, we find a good positive correlation between them. 

 In this study, we have performed our wind calculations in 2.5D, assuming axisymmetry. This allows us to perform a large number of simulations for a reasonable amount of computational resources. However, as evident from the discussion above, we know that non-axisymmetric components play a significant role at the maximum of the solar cycle. \cite{finl18} found that both axisymmetric and non-axisymmetric components field components impact the radial decay of magnetic flux in a similar pattern. \cite{gara16} also showed that torque generated by both axisymmetric and non-axisymmetric components are comparable. Motivated by these studies, we chose to alter the field strength for each $l$ mode by including the strength of both the axisymmetric and non-axisymmetric components following the quadrature addition formula:
 \begin{equation}
 \label{eq:brm}
     B_r^{\rm sim}(R_\odot,\theta) = \sum_\ell \left( \sum_{m=-\ell}^{\ell} \alpha_{\ell,m}^2\right)^{1/2} Y_{\ell,0}(\theta)\, .
 \end{equation}
where $\alpha_{\ell,m}$ is the strength for each $\ell$ and $m$ mode. One can see \cite{finl18a} for more details on this formulation.

We reconstructed the fields inside our PLUTO grid by combining these coefficients with spherical harmonics. Finally, we use the potential field source surface (PFSS) extrapolation \citep{scha69, alts69, schr03} method to extrapolate the magnetic fields inside our computational domain. We choose the source surface radius $R_{ss}$ as $100R_\odot$ for the PFSS extrapolation method. We use this extrapolated magnetic field as an initial magnetic field for our model that is then left to evolve self-consistently in the solar corona (see Appendix \ref{sec:appendixb}). 
 
\subsection{Boundary Conditions}
\label{sec:boundary}
\paragraph{Boundary Conditions for the Polytropic setup}
In the polytropic setup, we set an outflow boundary condition at the top boundary for all variables except the radial magnetic field; while we fix an axisymmetric boundary condition at the latitudinal boundaries ($\theta =0$ and $\pi$). We ensure the opening of magnetic field lines at the top boundary and the divergence free condition of the magnetic field by enforcing the condition $\frac{\partial}{\partial r}(r^2 B_r) =0$ at the top boundary. We set the conditions as prescribed by \cite{zanni09} for all variables at the bottom radial boundary.
 
\paragraph{Boundary Conditions for the Alfvén wave driven setup}
We use the same boundary conditions as the polytropic setup at the top and latitudinal boundary. We also set the outflow boundary condition at the top boundary and axisymmetric boundary condition at the latitudinal boundaries for the wave components ($\mathcal{E}^\pm$). The inner boundary condition is, however, different since the domain starts at the top of the transition region. We fix the plasma density, pressure, and the radial component of the magnetic field $B_r$. We want to model the star as a perfect conductor rotating with an angular speed $\Omega_{\odot}$. It implies that electric field should be zero in the reference frame co-rotating with the star. We enforce the zero electric field condition at the bottom boundary by setting $v_p \parallel B_p$. The pressure (or temperature) is modified dynamically to remain close to a hydrostatic equilibrium. Finally, the wave amplitude of the outward component is set to a fixed value $\delta z^{+} = 2 \, \delta v$, where $\delta v$ is the input transverse velocity. The inward component of the waves is set to 0. 

The resulting Poynting flux through our lower boundary is
\begin{equation}
    \label{eq:poyfdef}
    F_p = \langle \rho_\odot v_{A, \odot} \delta v_\odot^2 \rangle \approx \rho_\odot \langle v_{A,\odot} \rangle \delta v_\odot^2 \, . 
\end{equation}
It acts as an energy input to the solar wind. The input Poynting flux is chosen to remain near $10^5$~erg.cm$^{-2}$ s$^{-1}$. An additional heating term is also added, with a heating flux $F_h$ and a scale height of one solar radius, which acts essentially in the transition region and the low corona. We have varied the value of $F_h$ and $\delta v_\odot$ for the parameter space study described in Section \ref{sec:dependence}. The total input source flux in our model remains around $1.5 \times 10^{5}$ erg.cm$^{-2}$ s$^{-1}$, which corresponds to the energy flux of the solar wind in the outer heliosphere \citep[see][]{revi18}.

\begin{figure}[!ht]
\centering
\begin{tabular}{cc}
\includegraphics*[width=\linewidth]{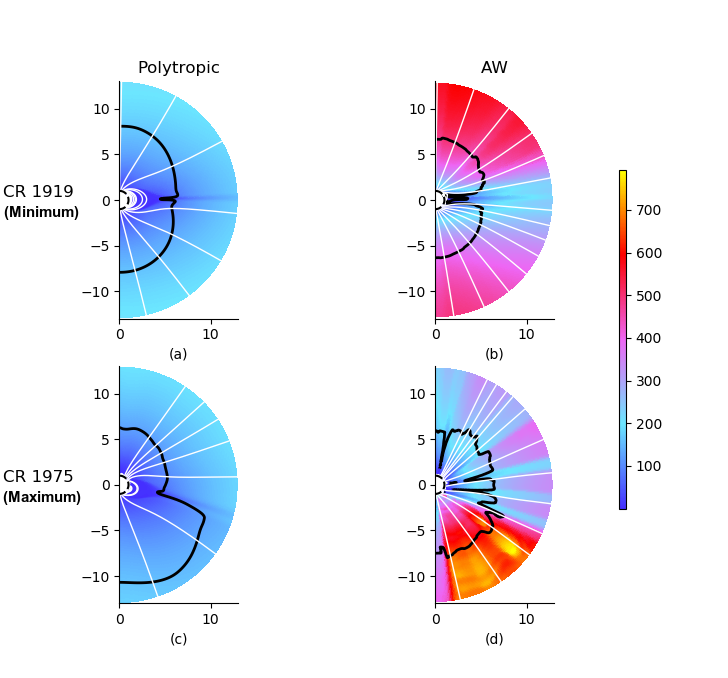}
\end{tabular}                
\caption{\footnotesize{Wind simulations considering the contribution of both axisymmetric and non-axisymmetric components by equation \eqref{eq:brm}. The top two panels show the poloidal velocity in the meridional plane for CR 1919 (cycle minimum) obtained from the (a) polytropic and (b) Alfvén wave driven solar wind models respectively, while the bottom two panels show the poloidal velocity in the meridional plane for CR 1975 (cycle maximum) obtained from the (c) polytropic and (d) Alfvén wave driven models. In each plot, the white solid lines indicate the magnetic field lines in the meridional plane, and the black line corresponds to the Alfvén surface. The poloidal velocity (color scale) is in km/s. This plot suggests that poloidal velocity is higher in Alfvén wave driven scenario compared to the polytropic, albeit the simulations achieve a similar mass loss.}}
\label{fig:alfv1}
\end{figure}

\section{Comparison between Polytropic Solar Wind Model and Alfvén Wave driven Solar Wind Model} 
\label{sec:comparison}

In this section, we aim to compare the results obtained from a polytropic solar wind model and an AW model. We have selected CR 1975 (near the maximum of the solar cycle 23) and CR 1919 (near the minimum of the solar cycle 23) for this purpose. We initialize both polytropic and AW simulations with the decomposed surface magnetic field data obtained from the synoptic maps. We have discussed the significant impact of non-axisymmetric components on the wind solutions in detail in Appendix \ref{sec:appendixa}. Motivated by the results described in Appendix \ref{sec:appendixa}, we choose to initialize the model considering the contribution of both axisymmetric and non-axisymmetric components following equation~\eqref{eq:brm}. One may see Appendix \ref{sec:appendixb} for the initial magnetic field topologies obtained from these two maps (Figure \ref{fig:snapshot-initial}). We have now four cases for the purpose of comparison: CR 1919 (polytropic and AW cases) and CR 1975 (polytropic and AW cases). We have denoted these four cases as 1919mP, 1919mAW, 1975mP, and 1975mAW for the rest of our paper.

\begin{figure}[!ht]
\centering
\begin{tabular}{cc}
\includegraphics*[width=\linewidth]{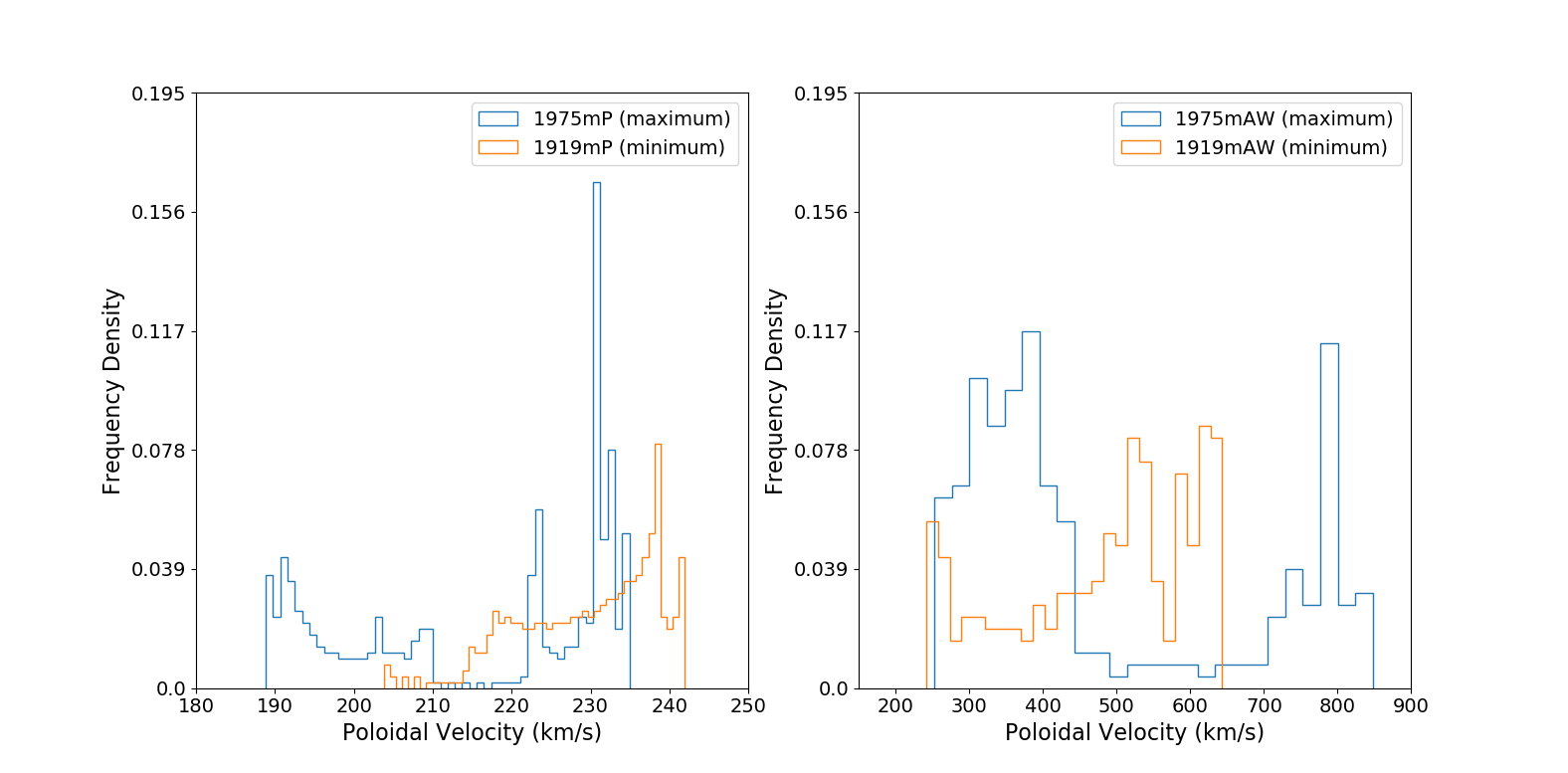}
\end{tabular}                
\caption{\footnotesize{Histograms of the poloidal velocity distribution at $20 R_\odot$ ($\approx$ 0.1 AU) for CR 1975 (blue) and CR 1919 (orange), for the polytropic model (left panel) and for the Alfvén wave driven model (right panel). The Alfvén wave driven wind model shows the bimodal distribution, while the polytropic model does not. }}
\label{fig:hist2}
\end{figure}

As the value of the mass loss rate is better known from the observations, we choose to compare the polytropic and the Alfvén wave driven wind model when both of the models will give almost similar mass loss value. We define the mass loss rate as: 
\begin{equation}
\dot{M} = \int \rho \, v_r d{S},
\label{eq:ml}
\end{equation}
where $\rho$ is the density and $v_r$ is the radial wind speed. We compute this integral over a spherical surface $S$ beyond the largest closed magnetic loop. Regardless the choice of integration surface, one will get the constant mass loss value as long as the previous condition is satisfied. Mass loss rate obtained from the polytropic solar wind model for CR 1919 and CR 1975 (1919mP and 1975mP cases) is $2.67 \times 10^{-14} \ M_\odot/ \textrm{yr}$ and  $2.75 \times 10^{-14} \ M_\odot/\textrm{yr}$ respectively. The mass loss rate in the AW model is $2.38 \times 10^{-14} M_\odot/\textrm{yr}$ for CR 1919 and $2.68 \times 10^{-14} M_\odot/\textrm{yr}$ for CR 1975 (1919mAW and 1975mAW cases). Mass loss rates are almost similar for all cases. One may see Table \ref{tab:massloss} in the Appendix \ref{sec:appendixa} which summarizes the value of mass loss in different scenarios. Despite this common global property, we nevertheless find significant differences between the results obtained from the polytropic and Alfvén wave driven solar wind model for both CR 1975 and 1919.

\subsection{On the wind speed bimodality}
\label{sec:wind-speed-bimod}

Figure \ref{fig:alfv1} (a) and (b) show the poloidal velocity in the meridional plane obtained from the polytropic and Alfvén wave driven solar wind models initialized with the CR 1919 WSO map (cycle minimum). Figure \ref{fig:alfv1} (c) and (d) show the same when the models are initialized with the WSO map of CR 1975 (cycle maximum). One will be able to understand the impact of solar wind dynamical pressure by comparing these steady state solutions (Figure \ref{fig:alfv1}) with their corresponding initial magnetic field line topology (Appendix \ref{sec:appendixb}). Initial magnetic field lines become open due to the dynamical pressure of the solar wind. When the solution achieves the steady-state, we find closed magnetic field lines at latitudes where we have the zero radial magnetic field. These regions are known as dead zones (or streamers).  By comparing the steady-state solutions at the maximum (CR 1975) and minimum (CR 1919) phases of the solar cycle, we find that the number of dead zones is controlled by the initial magnetic field topology. In summary,  there is a strong influence of the initial magnetic field topology on the solar wind structure. 

Figure \ref{fig:alfv1} indicates that we can obtain higher poloidal velocity in the Alfvén wave driven wind model compared to the polytropic model. We are able to reproduce the observed bimodality in the Alfvén wave driven wind model scenario. However, we are not able to reproduce the same in the polytropic wind model scenario. Poloidal velocity distribution is much wider in the Alfvén wave driven scenario compared to polytropic. Earlier studies indicated the importance of large scale energy deposition beyond the critical point in the supersonic regime to explain the observed velocity distribution \citep{leer80, leer82, stew97}. Since magnetic field lines are mostly open at 0.1 AU, we have also estimated the open flux from all of our models:
\begin{equation}
    \Phi_\textrm{open}= \int |B_r| d{  S}
\label{eq:of}    
\end{equation}
where $B_r$ is the radial magnetic field. Here also we compute the integral over a spherical surface beyond the largest closed magnetic loop. The magnetic flux becomes constant beyond a certain distance as field lines become open, and this constant value is the open flux. The open flux obtained from the polytropic solar wind model is $1.37 \times 10^{22}$\,Mx for CR 1919 and $1.36 \times 10^{22}$\,Mx for CR 1975. On the other hand, open flux obtained from the Alfvén wave driven solar wind model is $2.81 \times 10^{22}$\,Mx for CR 1919 and $3.29 \times 10^{22}$\,Mx for CR 1975. The open flux is therefore higher for the AW model compared to the polytropic one, and we also find that the open flux is higher at cycle maximum (CR 1975) than at cycle minimum (CR 1919) in the AW model.This is not the case in the polytropic model, that shows almost no variation in open flux between the two CRs. 

Next, we focus on the latitudinal distribution of the wind speed. We find slower wind around the dead zones (streamers) compared to the open magnetic field regime. Streamers are generally located at latitudes where magnetic forces are strong enough to counteract the thermal and dynamic pressure. Away from the streamer, we find the flux tubes that allow the wind to reach its maximum speed. On top of the dead zone, the magnetic field reaches its minimum value (zero in ideal MHD scenario). It is thus expected that the Alfvén surface, the surface at which the speed of the solar wind is equal to the local Alfvén speed ($v_A = \sqrt{||\boldsymbol{B}||/(\mu_0\rho)}$), will touch the top of the dead zone. Slow wind around the dead zones extends the Alfvén surface slightly on both sides from this minimum of the Alfvén surface. In these extended regions (both sides from the minimum of the Alfvén surface) thin current sheet structures are formed.  At the solar minimum (CR 1919), we find streamers at the low latitudes (near the equator) and open field line structures at high latitudes (see Figure~\ref{fig:alfv1}(a) and (b)).  Because of this, we mostly find the slow wind at the low latitude (near the equator) and fast wind at the high latitude (mostly from the polar coronal hole regime) during the solar minimum.  On the other hand, we find a complex coronal structure at the maximum of solar activity (CR 1975), with streamers distributed across different latitudes (see Figure~\ref{fig:alfv1} (c) and (d)). The only difference in the coronal structure between the cycle maximum and minimum is that streamers are differently located in latitude. We find that streamer size is bigger in the polytropic wind scenario compared to Alfvén driven one (see Figure \ref{fig:alfv1}). We also notice a drop in the wind speed at the high latitude at the cycle maximum due to the presence of closed magnetic loop structures at high latitudes. This kind of scenario is also found in the Ulysses data and has been interpreted as a high-latitude conic current sheet in the solar wind \citep{khab17}.

Let us now compare the profile of the Alfvén surface obtained at both activity maximum and minimum. As we already discussed, Alfvén surface in all of our four cases touches the top of the dead zone; when there is a current sheet formation (see Figure~\ref{fig:alfv1}). As the number of dead zones is higher at the cycle maximum (due to more complex topology), we find the Alfvén surface more irregular at the activity maximum compared to the activity minimum (see Figure~\ref{fig:alfv1} (d)). In the steady state, all closed magnetic field lines reside within the Alfvén radius. We find almost a similar geometrical average Alfvén radius for both the Alfvén wave driven and polytropic wind model setup ($5 R_\odot$ and $5.5 R_\odot$ respectively). Interestingly, we find a large asymmetry in the Alfvén radius for the 1975mP case. Alfvén radius is higher in the southern hemisphere compared to northern (see Figure~\ref{fig:alfv1} (c)). However, we do not find the same in the 1975mAW case. This is because there is an increase in both the poloidal velocity and Alfvén velocity in the Alfvén wave driven scenario. 

Finally, we study the distribution of the poloidal velocity amplitude at 20 solar radii (0.1\,AU) obtained from our four selected  models as shown in Figure~\ref{fig:hist2}. We confirm here again that the AW model (right panel) reaches higher wind speeds than the polytropic one (left panel). The polytropic model shows hints of a bi-modal distribution spanning a narrow range of about 100\,km/s, and reaching a maximum 280\,km/s. Conversely, the AW model strikingly exhibit a solar-like bi-modal distribution, especially at solar maximum, with a slow component near 350 km/s and a fast component close to 800\,km/s. Note that the solar wind speed usually lies between 400 and 900\,km/s at 1\,AU. Near the solar minimum (CR 1919 case), the AW model achieves maximum wind speeds up to 650\,km/s at 0.1\,AU. In summary, we are able to better reproduce observed bimodality and distribution breadth of wind speeds with the AW model.

\begin{figure}[!ht]
\centering
\includegraphics*[width=0.5\textwidth]{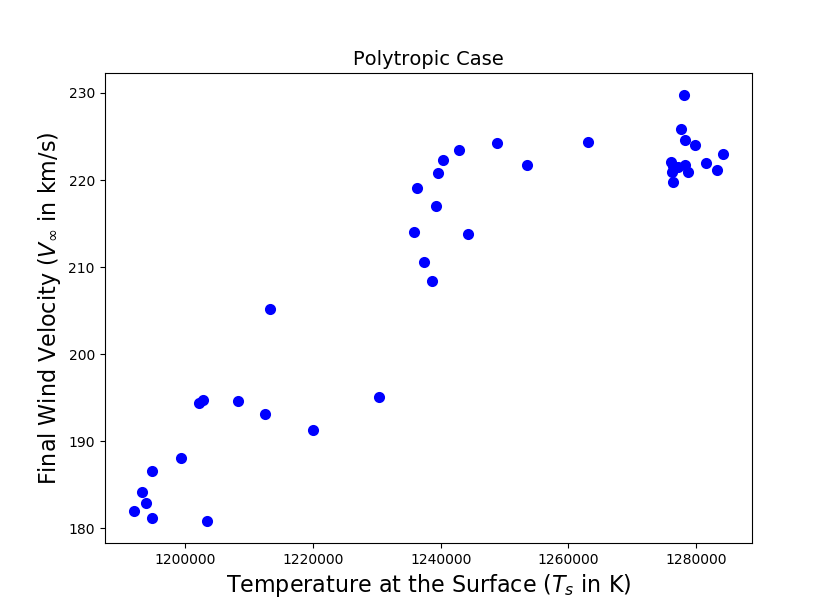}\\
\caption{\footnotesize{Relationship between the final wind velocity ($V_\infty$) and the surface temperature ($T_s$) obtained from the polytropic solar wind model. The Spearman correlation coefficient is 0.86.}}
\label{fig:veltemp1}
\end{figure}

\begin{figure}[!ht]
\centering
\begin{tabular}{cc}
\includegraphics*[width=\linewidth]{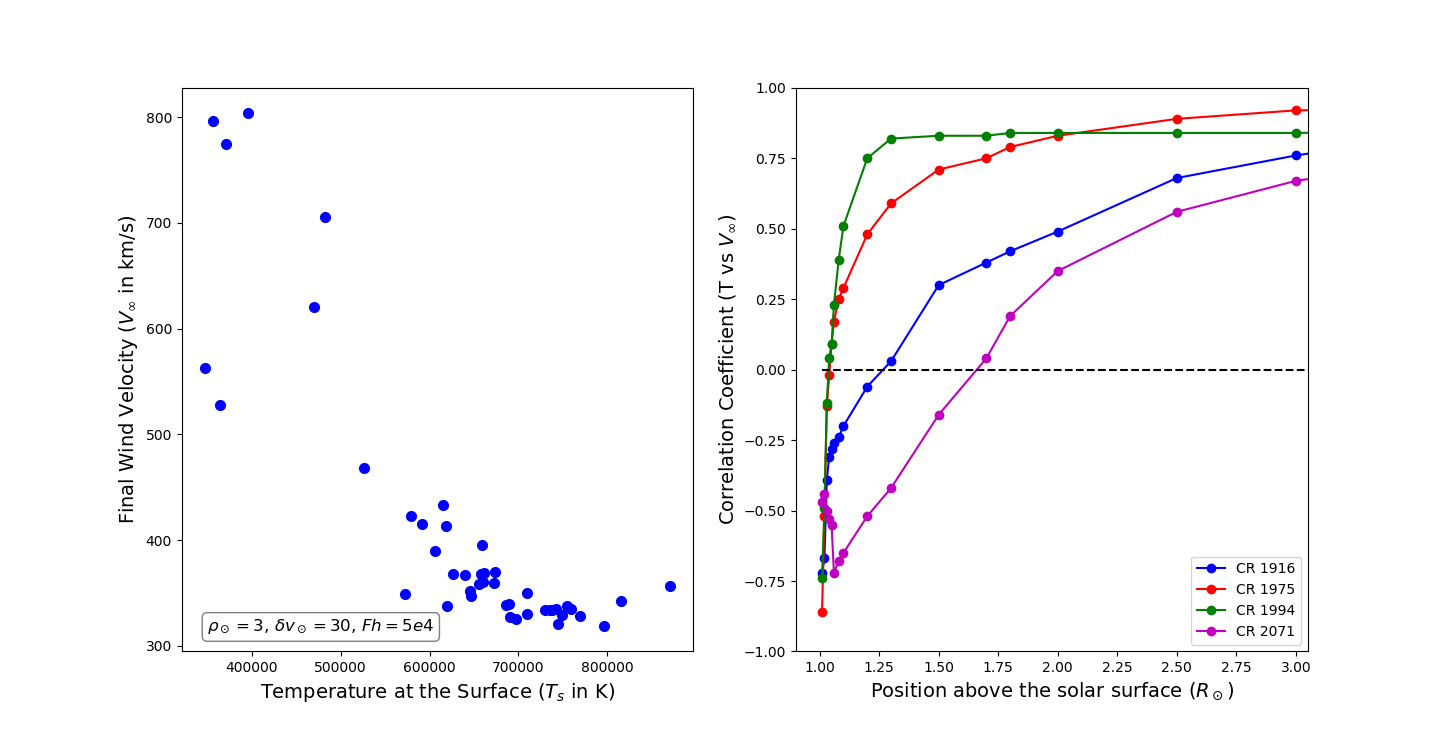}\\
\end{tabular}                
\caption{\footnotesize{The left panel shows the relationship between the final wind velocity $V_\infty$ and the surface temperature $T_s$ obtained from the Alfvén wave driven solar wind model. The Spearman correlation coefficient is -0.89. The right panel shows how the Spearman correlation coefficient between the final wind velocity $V_\infty$ and temperature $T$ changes if we consider the temperature at different positions of the solar corona. The black dashed line (line of no correlation) indicates the transition from negative correlation to positive correlation. The colors correspond to different magnetic maps. We notice that transition takes place at a larger distance from the solar surface at the cycle minimum than at the cycle maximum. Please note that we use the same initial parameters, namely, coronal base density $\rho_\odot = 3$ in units of $10^8$\,cm$^{-3}$, input transverse velocity 30\,km/s, and heating flux $5 \times 10^4$\,erg\,cm$^{-2}$\,s$^{-1}$, as mentioned in the legend of the left panel.}}
\label{fig:veltemp2}
\end{figure}

\subsection{On the wind speed-source temperature relationships}
\label{sec:wind-speed-T}

One important observational feature is the anti-correlation between the final solar wind speed and the corresponding coronal source temperature \citep{geis95}. Many studies have tried to explain this observed anti-correlation. \cite{schw03} explain the observed anti-correlation as a consequence of strong temperature dependence of the thermal conduction term. In summary, slow wind loses more energy from the thermal conduction due to higher coronal temperature. One may also see \cite{schw11, schw14} for the details regarding this. However, some other studies with the AW models indicate that higher electron temperature at the source of the slow solar wind may cause flux tubes with different expansion factors to have different Parker critical points and different radial distribution of heating rate \citep{leer80, cran05a, cran10, cran10a, cran17}. Here we want to study the relationship between the terminal wind velocity and the coronal source temperature from our wind models.

We have already shown that both of our wind models can generate faster and slower solar wind streams but differ by the amplitude difference between the fast and slow wind streams (see Figure \ref{fig:hist2}). We now want to figure out the relationship between the final wind velocity and the coronal source temperature from the polytropic wind model. As the solar wind accelerates along the open field lines, we consider the temperature at the beginning point of an open field line (near the coronal base) as source temperature ($T_s$), and the wind velocity at the endpoint of the open field line (near 20 solar radii) as the final wind velocity ($V_\infty$). Please note that the endpoint of the field line is not the physical endpoint of the field line but the computational one. We find a positive correlation (Spearman correlation coefficient 0.86) between the final wind velocity and the coronal source temperature from the polytropic wind model (see Figure \ref{fig:veltemp1}). Sometimes we also notice a weaker positive correlation for some magnetic maps. The polytropic (thermally driven) wind model always shows a positive correlation between the final wind velocity and the temperature, even if we consider the temperature away from our lower coronal boundary. Previous studies based on the polytropic assumption also suggest a positive correlation between the final wind velocity and the coronal source temperature \citep{leer82}. This kind of relationship is also expected in the Parker wind scenario. However, observations show the opposite. In summary, we are not able to explain the observed anti-correlation between the final wind velocity and the source temperature from the polytropic solar wind model.

Next, we aim to characterize this same relationship using our AW model. The left panel of Figure \ref{fig:veltemp2} shows an anti-correlation (correlation coefficient -0.89) between the final solar wind speed ($V_\infty$) and the coronal source temperature ($T_s$). Final solar wind speed decreases from 810\,km/s to 300\,km/s with the increase of surface temperature. Our AW model is therefore able to reproduce the observed anti-correlation between the final wind velocity and the coronal source temperature. Probably different radial distribution of heating rate or the strong temperature dependence of thermal conduction is responsible for this result in our model \citep{schw03, cran05}. However, if we consider the temperature away from the bottom boundary, after some distance we get a positive correlation between the final wind velocity and the temperature (see right panel of Figure \ref{fig:veltemp2}). 

To illustrate this point, we chose two magnetic maps near the maximum of the solar cycle 23 (CR 1975 and 1994) and two magnetic maps near the minimum of the solar cycle 23 (CR 1916 and 2071). The right panel of Figure \ref{fig:veltemp2} indicates that at the maximum of the solar cycle, the shifting from negative to positive correlation takes place very close to the solar surface (only at 1.05 solar radii). However, this is not the case at the minimum of the solar cycle. The shift from the negative correlation to positive correlation takes place at a comparatively larger distance (1.3 solar radii for CR 1916 and 1.8 solar radii for CR 2071). Shifting distance lies between 1.04 and 2 solar radii in our Alfvén wave driven setup. We did not identify precisely why the shift from negative to positive correlation occurs at different distances depending on the structure of the initial magnetic map. It could originate from  a decrease in the map complexity or in the total surface energy of the magnetic map. As the two are correlated (see the bottom panel of Figure \ref{fig:surfen}), additional simulations would be required to disentangle their relative effects, which is outside the scope of the present paper. In summary, AW model can well reproduce the fact that fast wind streams originate from the cool coronal holes, and slow wind streams from near streamers like structure. 

\newcommand*\mycolhead[1]{\multicolumn{1}{p{1.7cm}}{\raggedleft #1}}
\begin{longtable}[!ht]{*8r}
\caption{Values of the mass loss rate, of the angular momentum (A.M.) loss rate, and of the average Alfvén radius, for different combination of coronal parameters. }
\label{tab:list}\\
 \hline
 \multicolumn{8}{c}{Cycle Maximum CR 1975}\\[0.50ex]
 \hline
 \mycolhead{Base density $\rho_\odot$ ($10^8$\,cm$^{-3}$)} &
 \mycolhead{Transverse Velocity $\delta v_\odot$ (km/s)} &
 \mycolhead{Heating Flux $F_h$ ($10^4$ erg cm$^{-2}$ s$^{-1}$)} &
 \mycolhead{Poynting Flux $F_p$ ($10^4$ erg cm$^{-2}$ s$^{-1}$)} &
 \mycolhead{$\Upsilon_\textrm{open}/(4\pi)$} &
 \mycolhead{Mass Loss ($10^{-14}$ $M_\odot/\textrm{yr}$)} &
 \mycolhead{A.M. Loss ($10^{29}$ erg)} &
 \mycolhead{Alfvén Radius ($R_\odot$)}\\
  \hline
3	& 30	&0.05	& 3.73& 303.71 & 1.18	&3.31& 5.956 \\
3 &	30	&0.1&	3.72 &302.71 & 1.20&	3.34 & 5.948\\
1&	30	&2.0&	2.04& 338.11 & 0.85	&2.29 & 5.877 \\
3	&30&	0.54& 3.69 & 286.2 &	1.31&	3.47& 5.801 \\
0.5	&30	&2.0 & 1.48&423.46&	0.55&	1.71 & 6.278 \\
3	&30	&2.0  & 3.61& 240.73&	1.69&	3.84& 5.392 \\
0.3	&30	&2.0& 1.16& 507.85 &0.40	&1.37& 6.635\\
3	&10	&2.0& 0.44	&395.02&0.48	&1.99& 7.343 \\
0.3	&10	&2.0& 0.13	&737.53&0.19	&1.06& 8.408 \\
3	&30&10.0	& 3.38 & 136.29&3.80&	5.24& 4.186 \\
3 &	30	&5.0	& 3.51& 187.18 &2.4&	4.49& 4.831 \\
1	&30	&5.0	& 1.99&251.7&1.42	&3.04& 5.207 \\
1	&30	&10.0	& 1.97&176.73&2.44	&3.79 & 4.439 \\
1	&30	&0.1	& 2.09&532.64&0.49	&1.84 & 6.947 \\
1	&40	&2.0	& 3.55& 255.25& 1.36	&2.88& 5.183 \\
2 & 40 & 2.0  & 4.96 & 195.32 & 2.35 & 4.37& 5.168 \\
0.5 & 40 &2.0 & 2.58 & 343.72 & 0.78 & 1.87 & 5.898 \\
0.3 & 40 & 2.0& 2.04 & 433.76& 0.52 & 1.43& 6.296 \\
2 & 10 & 2.0 & 0.35& 508.39 & 0.95 & 2.24 & 7.948\\
1 & 10 & 2.0 & 0.24 & 634.9 & 0.35& 1.97& 8.452 \\
0.5 & 10 & 2.0 & 0.17 & 778.04 & 0.21 & 1.23 & 8.727 \\
1 & 30 & 2.0 & 2.03 & 338.11 & 0.85 & 2.29 & 5.877 \\
3 & 25 & 2.0 & 2.59& 284.15 & 1.23 & 3.28 & 5.825 \\
3 & 20 & 2.0 & 1.69 & 337.83 & 0.89 & 2.81 & 6.353 \\
3 & 15 & 2.0 & 0.97 & 313.78 & 0.61 & 1.85 & 6.253 \\
1 & 30 & 0.54 & 2.07 & 451.01 &  0.60 & 2.03 & 6.583 \\
1 & 10 & 2.0 & 0.24 & 634.9 & 0.36& 1.97 & 8.763 \\
1 & 15 & 2.0 & 0.53 & 563.01 & 0.42 & 2.01 & 7.769 \\
1 & 25 & 2.0 & 1.44 & 398.27 &0.67 & 2.16 & 6.396 \\
2 & 30 & 2.0 & 2.85 & 284.83 & 1.35 & 3.44 & 5.686 \\
\hline 
\multicolumn{8}{c}{Cycle Minimum CR 1919}\\[0.50ex]
 \hline
 
1&	30&	0.54 & 1.72& 490.83&	0.65&	2.34 & 6.775 \\
3	&30&	0.54 & 2.81 & 283.50&	1.15&	2.76 & 5.535 \\
2	&30&	2.0  & 2.29&	257.81 &1.33&	2.91 & 5.265 \\
2	&40&	2.0  & 4.17& 169.33 &	2.03&	2.91 & 4.304  \\
0.5	&40&	2.0  & 2.02 & 339.96&	1.02&   2.46 & 5.557 \\
0.3	&40&	2.0 & 1.58 &	444.66 &0.76&	2.27 & 6.149 \\
2	&10&	2.0  & 0.27&593.08&	0.56&	2.94 & 8.156 \\
1	&10&	2.0  & 0.19& 674.16&	0.52&	3.02 & 8.637 \\
1	&20&	2.0 & 0.73&	508.52&   0.69&	2.80& 7.204 \\
1	&15&	2.0  & 0.42&603.19&	0.57& 2.82& 7.940 \\
3	&15&	2.0  & 0.73&466.13&	0.72&	2.87 & 7.130 \\
3	&20&	2.0  & 1.30& 365.71& 0.94&	2.96& 7.519 \\
3	&25&	2.0  & 1.98& 287.46&	1.21&	2.96 & 5.601 \\
0.5	&10&	2.0  & 0.13& 774.22&	0.44&	2.76 & 8.979 \\
0.3	&10&	2.0 & 0.11&	799.41 & 0.31&	1.96 & 8.921 \\
1	&25&	2.0  & 1.12& 411.25&	0.85&	2.79 & 6.485 \\
3 & 10 &    2.0  & 0.32&  580.91& 0.56&   2.82 & 7.988 \\
1   &30&    2.0  & 1.61&  336.10& 1.03&   2.77 & 5.838 \\
0.5 &30&    2.0  & 1.14& 463.17 & 0.74&   2.51 & 6.563 \\
0.3 &30&  2.0  & 0.91&  564.57 & 0.59&   2.31 & 7.069 \\
3   & 30 &  2.0 & 2.79& 221.17& 1.57&  2.99 & 4.953 \\
1 & 10&   2.0  & 0.19& 674.16&  0.52&  3.02&  8.637 \\
3  & 30&    0.1  & 2.83&  305.44& 1.02&  2.48E&  5.625 \\
3  & 30&    0.05  & 2.83& 306.32&  0.98&  2.44&  5.627 \\
3   &30&    10.0  & 2.74& 97.37&  3.75&  3.32& 3.357 \\
3  &30&     5.0  & 2.79& 149.80&  2.38&  3.17& 4.118 \\
1  &30&     5.0  & 1.58& 199.50&  1.80&  3.0& 4.610 \\
1  &30&     10.0  & 1.55& 121.65&  3.01&  3.10& 3.629 \\
1  &30& 0.05  & 1.73& 591.87&  0.51&  2.06& 7.189 \\
1  &30&  0.1 & 1.73& 554.95& 0.54&  2.12& 7.055 \\
3 & 40& 2.0 & 5.01& 138.57 &2.45& 3.04 & 3.978 \\
1 & 40 & 2.0 & 2.83& 283.60 & 1.46& 2.73& 4.873 \\
\hline 
\end{longtable}

\section{Dependence of the solar wind properties on the coronal parameters in the Alfvén wave driven solar wind model}
\label{sec:dependence}

We now turn to the study of the influence of the coronal parameters of the AW model on the global properties of the modelled wind. We will characterize the global solution by three numbers: the mass loss rate, the angular momentum loss rate, and the Alfvén radius of the wind solution. We have already defined the mass loss rate in Eq.~\eqref{eq:ml}. Here, we will define the other two quantities, namely, the angular momentum loss rate and the average Alfvén radius.

In this study, we define the rate of angular momentum loss following the equation:
\begin{equation}
\dot{J} = \int_S \rho \, \Lambda  \, v_r\;d{S},
\label{eq:jl}
\end{equation}
where $\Lambda = R_0 \left(v_\phi - \frac{B_\phi}{\mu_0\rho} \frac{\boldsymbol{B}_p\cdot\boldsymbol{v}_p}{||\boldsymbol{v}_p||^2}\right)$.
Here $\boldsymbol{B}_p$ and $\boldsymbol{v}_p$ represents the poloidal magnetic field and poloidal velocity respectively. $R_0$ is the spherical radius where we want to calculate the angular momentum loss rate. This integral is computed over a spherical surface $S$ (with radius $R_0$) which contains all closed magnetic field structures. Value of the angular momentum loss does not depend on the value of $R_0$ as long as it contains all closed magnetic structures.

We can also describe the angular momentum loss rate in terms of Alfvén radius. \cite{webe67} first quantify the angular momentum carried by the plasma using their one dimensional model and find the specific (per unit mass) angular momentum ($I_w$) as:
\begin{equation}
    I_w= \Omega_* R_A^2
\label{eq:lw}    
\end{equation}
where $\Omega_*$ is the stellar rotation rate and $R_A$ is the Alfvén radius. One can then express the stellar angular momentum loss rate in a steady state as a product of the specific angular momentum ($I_w$) carried by the solar wind and the mass loss rate:
\begin{equation}
    \dot{J}= \dot{M} \Omega_* R_A^2
\label{eq:jr}    
\end{equation}
We then define the average Alfvén radius following Equation \eqref{eq:jr}:
\begin{equation}
    \langle R_A \rangle= \sqrt{\frac{\dot{J}}{\dot{M} \Omega_*}}
\label{eq:ra}    
\end{equation}
We have computed the angular momentum loss rate $\dot{J}$ and the mass loss rate $\dot{M}$ from our simulations when the solution have reached a steady state. We then calculated the average Alfvén radius following Equation \eqref{eq:ra}.

 \begin{figure*}[!ht]
\centering
\begin{tabular}{cc}
\includegraphics*[width=\linewidth]{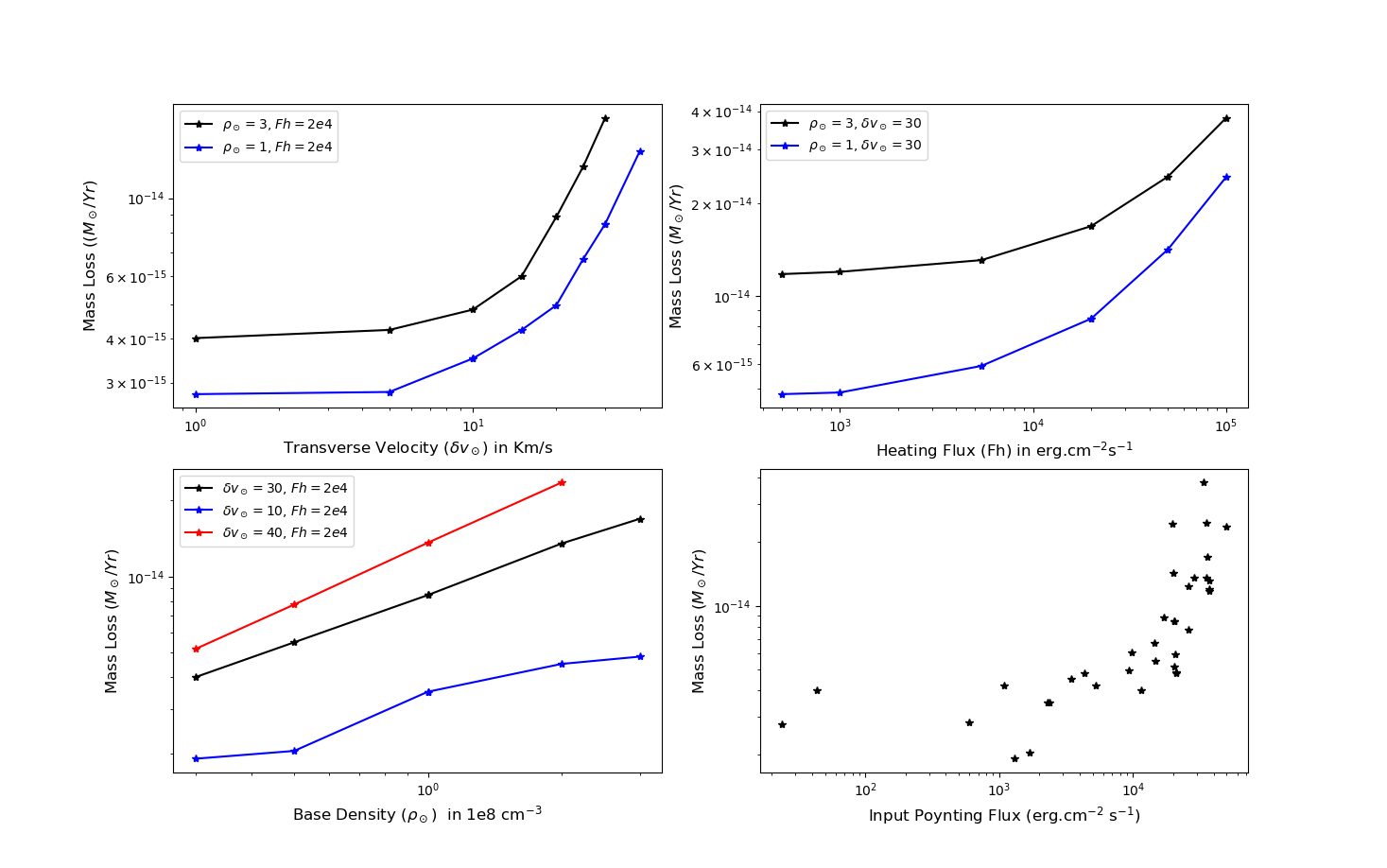}\\
\end{tabular}                
\caption{\footnotesize{Variation of the mass loss with different coronal parameters: input transverse velocity $\delta v_\odot$, input coronal heating flux $F_h$, input base density of the corona $\rho_\odot$, and input Poynting flux $F_p$. We initialize these Alfvén wave driven wind simulations with the WSO magnetic map of CR 1975. Lines with different colors (blue, black and red) correspond to different combination of fixed coronal parameters. We mention these combinations in the legend of the figures.}}
\label{fig:dep1}
\end{figure*}

\begin{figure}[!ht]
\centering
\begin{tabular}{cc}
\includegraphics*[width=1.05\linewidth]{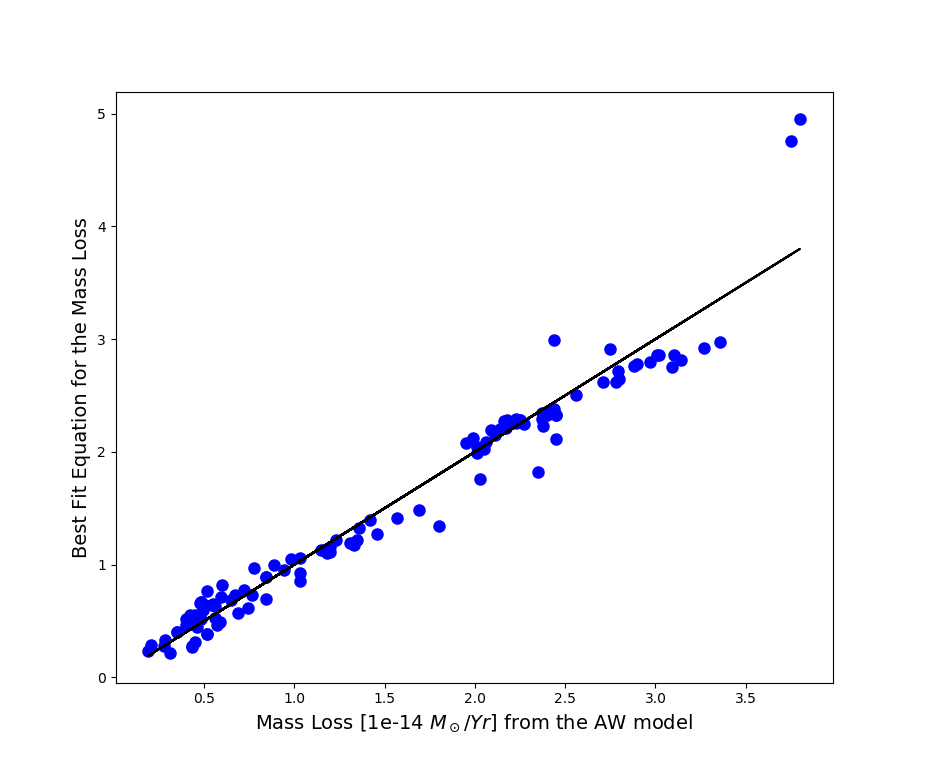}\\
\end{tabular}                
\caption{\footnotesize{Single law fit for the mass loss obtained from the Alfvén wave driven solar wind model, following Equation \eqref{eq:scaling-massloss}.}}
\label{fig:scaling}
\end{figure}

\subsection{General solar wind properties with respect to different sources of heating}
\label{sec:GeneralTrends}

The total energy input to solar wind in the Alfvén wave driven solar wind model is basically controlled by the amount of input Poynting flux and the coronal heating flux $F_h$ at the base of the solar corona. The total input Poynting flux $F_p=\rho_\odot \langle v_{A,\odot} \rangle \delta v_\odot^2$ at the coronal base in the AW model depends on the input density at the base of the corona $\rho_\odot$, on the input transverse velocity ($\delta v_\odot$), and on the initial magnetic field structure. We have assessed the relative influence of these parameters in a systematic exploration. We report all the models we explored in Table~\ref{tab:list}, along which we list the obtained rate of mass loss, the rate of angular momentum loss (A.M. loss), and the Alfvén radius. We have performed this systematic exploration for minimum and maximum phases of solar cycle 23 using the magnetic maps for CR 1975 and CR 1919. 

Figure \ref{fig:dep1} shows the variation of the mass loss with the different coronal parameters for an initial magnetic map (CR 1975) taken at the maximum of solar activity. The top left panel of Figure \ref{fig:dep1} shows an exponential increase in the mass loss when we increase the input transverse velocity $\delta v_\odot$  from $2$\,km/s to $40$\,km/s, while keeping the other two parameters, namely, the input coronal heating flux $F_h$ and the base coronal density $\rho_\odot$ fixed. The mass loss rate also increases exponentially with the input coronal heating flux $F_h$ as seen in the top right panel. Conversely, we find in the bottom left panel a power law increase with a slope $\sim 0.5$, in the rate of mass loss with the change of coronal base density $\rho_\odot$. The sublinear dependency of the mass loss on the base density can be understood by the increase of radiative losses with the density (see equation \ref{eq:radloss}). In the bottom right panel, we find that the mass loss dependency to the total Poynting Flux is exponential. This exponential behaviour of $\dot{M}$ with $F_h$ and $F_p$ is the result of the complex dynamics of the low corona inner boundary where the temperature evolves dynamically with the balance of heating terms. This is in agreement with earlier results which suggested that deposition of more energy at the coronal base leads to higher mass loss \citep{leer80}. We also find a similar kind of mass loss variation with different coronal parameters if we initialize our solar wind solution with the magnetic map CR 1919 taken at the solar minimum. Please note that we have not considered the contribution of the non-axisymmetric components in this calculation, but we expect them to play a negligible role in the parametrization of the mass loss here.


\subsection{Mass loss rate scaling law in the AW model}
\label{sec:MdotLawAW}

We have seen that the deposition of energies at the base of the solar corona in the AW model comes from the input coronal heating flux ($F_h$) and other initial parameters namely, density ($\rho_\odot$), initial field strength ($B_\star$) and the input transverse velocity ($\delta v$). This realization allows us to find a single scaling relationship between mass loss $\dot{M}$ and input energy parameters. We are able to fit all our simulations listed in Tables \ref{tab:list} and \ref{tab:list1} into one single law as shown in Figure \ref{fig:scaling}.
\begin{equation}
    \ln \dot{M} = a + b\,F_h + c\,\ln \rho + d\,\ln B_\star + e\,\delta v
\label{eq:scaling-massloss}
\end{equation}
The numerical values of the parameters of this scaling law are: $a=-2.23 \pm 0.13$, $b=1.53 e^{-5} \pm 8.29e^{-7}$, $c= 0.46 \pm 0.024$,  $d= 0.19 \pm 0.04$, and  $e= 0.0403 \pm 0.001$. If one uses the value of heating flux ($F_h$) in erg.cm$^{-2}$.s$^{-1}$, density ($\rho$) in $10^{8}$ cm$^{-3}$, transverse velocity ($\delta v$) in km/s and $B_\star$ in Gauss, then one will get the value of mass loss in units of $10^{-14}$ $M_\odot/\textrm{yr}$ from the scaling law. 
 
 
 Our best fit formulation indicates that mass loss is positively correlated with the magnetic field strength $B_\star$. However, \cite{revi15} used the polytropic solar wind model and found that mass loss is weakly dependent (negative correlation) on the magnetic field strength but increases strongly with the rotation rate. Our model predicts exactly the opposite about the magnetic field dependency that one can expect from the polytropic wind model. In the case of their fixed temperature polytropic solar wind model, mass loss rate reduces with the strength of the magnetic field as more close magnetic loops are formed in their model which confines the plasma. On the other hand, the energy source term responsible for wind driving changes with the magnetic field strength in the AW model. That's why mass loss is positively correlated with the magnetic field strength in our model. \cite{cohe14} also found a similar trend with their AW model. Finally, note that we perform all our wind simulations with a fixed rotation rate. Because of this, our scaling law does not have any dependency on the rotation rate.
 
 \subsection{Alfvén radius scaling law in the AW model}
\label{sec:RaLawAW}

 In this paragraph, we derive a scaling law for the Alfvén radius depending on the value of the open flux instead of the surface magnetic field strength. We recall that we use here an average Alfvén radius defined in Eq.~\eqref{eq:ra}. As the mass loss and the angular momentum loss occur through the open field lines, it is expected that Alfvén radius and the open flux are closely related. 
 
 We follow the topology-independent formulation developed by \cite{revi15,revi16} to derive the scaling law for the average Alfvén radius:
 \begin{equation}
     \frac{\langle R_A \rangle}{R_*}= K_1 \left[ \frac{\Upsilon_\textrm{open}}{\left(1 + f^2 / K_2^2 \right)^\frac{1}{2} }\right]^m
\label{eq:scaling-alfv}
 \end{equation}
 where $\Upsilon_\textrm{open}$ is a magnetization parameter which takes the value of open flux into account:
 \begin{equation}
     \Upsilon_\textrm{open} =\frac{\Phi_\textrm{open}^2}{R_*^2 \dot{M} v_\textrm{esc}}
\label{eq:upsilon}     
 \end{equation}
where, $R_*$ is the solar (stellar) radius, $\dot{M}$ is the mass loss and $v_\textrm{esc}= \sqrt{(2 G_{\odot} M_{\odot})/R_{\odot}}$ is the escape velocity.  $\Upsilon_\textrm{open}$ carries the information regarding the open field region and dead zones. It actually quantifies the value of open flux in our simulation compared to the surface magnetic flux. In Equation~\eqref{eq:scaling-alfv}, $K_1$, $K_2$ are the constant and $f$ is the fraction of stellar break up rate. One may also see \cite{matt08, matt12} for this kind of formulation. The fraction of the break up rate is defined as the ratio between the stellar rotation rate at the equator and the Keplerian speed; that is represented by the formula:
\begin{equation}
    f=\frac{\Omega_{\odot} R_{\odot}^\frac{3}{2}}{\sqrt{G M_{\odot}}}
\end{equation}
In this study, all of our simulations are performed with the constant rotation rate $\Omega_\odot$ that means fraction of the break up rate $f$ is constant for this study. In our all simulations, we use the value of $f$ as 0.004 (solar value). As there is no variation in the value of break up rate $f$ in our simulations, it is not necessary to fit the value of $K_2$ in Equation~\eqref{eq:scaling-alfv}. Moreover, \citet{revi15,revi16} found a value $K_2 = 0.06$, and $\sqrt{1+f^2/K_2^2} \sim 1.0022$. In the following, we thus assume the denominator of the right hand side of equation \ref{eq:scaling-alfv} is unity.

\begin{longtable}[!ht]{*7r}
\caption{Value of the mass loss rate, of the angular momentum (A.M.) loss rate, and of the average Alfvén radius obtained from the Alfvén wave driven wind simulations initialized with the different Carrington rotation maps over cycle 23. }
\label{tab:list1}\\
 \hline
 \multicolumn{7}{c}{Alfvén wave driven wind model}\\[0.50ex]
  \multicolumn{7}{c}{(Both Axisymmetric and Non-axisymmetric component)}\\[0.50ex]
 \hline
 Carrington Rotation & Fractional Year & Poynting Flux ($F_p$) & $\Upsilon_\textrm{open}/(4\pi)$ & Mass Loss & A.M. Loss & Alfvén Radius \\ [0.50ex]
 (CR) & (yr) & ($10^4$ erg.cm$^{-2}$ s$^{-1}$) & &($10^
 {-14}$ $M_\odot/\textrm{yr}$) & ($10^{29}$ erg) & (${\rm R_\odot}$)  \\ 
 
 \hline
	1916 & 1996.86 & 3.04 & 120.85&2.19&	2.60 & 3.833 \\
	1927 & 1997.68& 6.56 & 98.37 &2.71 & 3.10& 3.805\\
	1941 & 1998.73& 6.57 & 51.17& 2.78 & 2.42 & 3.324 \\
	1951 & 1999.48& 9.52 & 43.01 & 3.14 & 2.8 & 3.363\\
	1953 & 1999.63& 11.52 & 174.88 & 2.75 & 4.82& 4.712\\
	1961& 2000.23& 9.31 & 77.51& 2.97 & 3.59 & 3.919\\
	1965 & 2000.52& 8.59 & 79.76 & 2.88 & 3.66 & 4.017\\
	1970& 2000.89& 10.38 & 78.47 & 3.10 & 3.69 & 3.886\\
	1975 & 2001.27& 9.01 & 178.73 & 2.90& 4.91 & 4.638\\
	1980 & 2001.64& 11.62 & 136.91 & 3.27& 4.79& 4.317\\
	1985 & 2002.02& 12.87& 112.78 & 3.36 & 4.69& 4.210\\
	1989 & 2002.32& 8.57 & 81.05 & 3.09& 3.41 & 3.748\\
	1994 & 2002.69& 10.45 & 99.45& 3.02 & 4.31 & 4.265\\
	2005 & 2003.51& 6.87 & 131.96& 2.80 & 3.80 & 4.159\\
	2019 & 2004.55& 7.93 & 55.46 & 2.79 & 2.34 & 3.260\\
	2032 & 2005.52& 5.17 & 66.34 &2.56 & 2.41 & 3.456\\
	2043 & 2006.35& 3.27 & 98.06 & 2.23 & 2.48 & 3.767\\
	2058 & 2007.47& 3.14 & 100.63& 2.16& 2.52 & 3.848\\
	2071 & 2008.47& 1.71 & 56.69& 2.05 & 1.39 & 2.948\\
 \hline 
 \multicolumn{7}{c}{Alfvén wave driven wind model}\\[0.50ex]
 \multicolumn{7}{c}{(Only Axisymmetric Component)}\\[0.50ex]
 \hline
 Carrington Rotation & Fractional Year& Poynting Flux ($F_p$) & $\Upsilon_\textrm{open}/(4 \pi)$ & Mass Loss & A.M. Loss & Alfvén Radius \\ [0.50ex]
 (CR) & (yr)&($10^4$ erg.cm$^{-2}$ s$^{-1}$) & &($10^
 {-14}$ $M_\odot/\textrm{yr}$) & ($10^{29}$ erg) & (${\rm R_\odot}$)  \\ 
 \hline
 1916 & 1996.86& 2.33 & 113.54& 2.13 & 2.37 & 3.760\\
 1927 & 1997.68& 3.21 & 127.01& 2.25 & 2.72 & 3.921\\
 1941 & 1998.73& 3.18 & 72.06 & 2.18 & 2.14 & 3.533\\
 1951 & 1999.48&3.15 & 60.88& 2.25 & 2.14 & 3.478\\
 1953 & 1999.63& 3.98 & 72.39&  2.44 & 2.63 & 3.709\\
 1961 & 2000.23&2.93 & 79.23 & 2.27 & 2.52 & 3.756\\
 1965 & 2000.52&2.96 & 80.09 & 2.23 & 2.42 & 3.713\\
 1970 & 2000.89 &2.67 & 71.34 & 2.17 & 2.14 & 3.545\\
 1975 & 2001.27 &3.51 & 187.86& 2.45 & 4.49 & 4.834\\
 1980 & 2001.64 &3.26 & 37.22 & 2.37 & 3.21 & 4.146\\
 1985 & 2002.02 &3.72 & 111.87& 2.41 & 3.25 & 4.145\\
 1989 & 2002.32 &2.34 & 81.15& 2.11 & 2.11 & 3.573\\
 1994 & 2002.69 &3.64 & 90.73 & 2.37 & 2.91 & 3.961\\
 2005 & 2003.51 &2.62 & 91.38 & 2.14 & 2.39 & 3.772\\
 2019 & 2004.55 &2.17 & 60.15& 1.99 & 1.54 & 3.141\\
 2032 & 2005.52 &1.77 & 59.69 & 2.01 & 1.54 & 3.116\\
 2043 & 2006.35 &1.96 & 58.24 & 2.06 & 1.36 & 2.901\\
 2058 & 2007.47 &1.94 & 59.05 & 1.95 & 1.36 & 2.985\\
 2071 & 2008.47 &1.55 & 48.98 & 2.01 & 1.26 & 2.829\\
 \hline
\end{longtable}

\begin{figure}[!ht]
\centering
\begin{tabular}{cc}
\includegraphics*[width=1.0\linewidth]{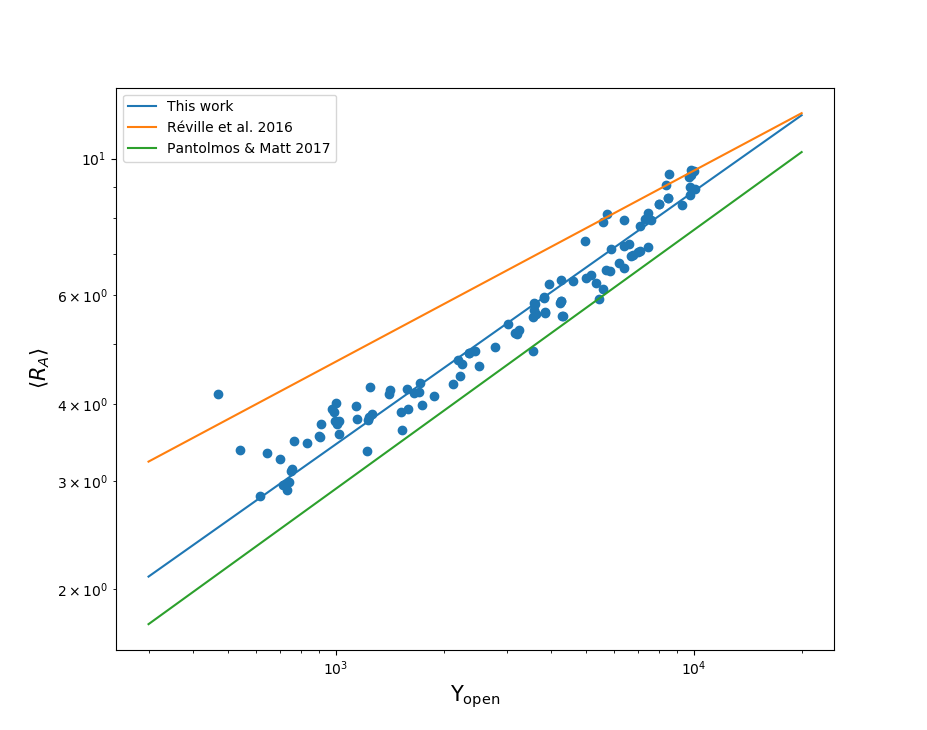}\\
\end{tabular}                
\caption{Single law fit for the average Alfvén radius obtained from the Alfvén wave driven solar wind model (blue dots) according to equation \eqref{eq:scaling-alfv}. Best fitted parameters for the Alfvén wave driven (AW) model are: $K_1= 0.20 \pm 0.004$ and $m=0.41 \pm 0.002$. The scaling law of \citet{revi16} and \citet{pant17} is shown in orange and green respectively. We see that despite different fitted parameters, the results of the AW model are compatible with the law of \citet{revi16} and \citet{pant17}, obtained with polytropic simulations.}
\label{fig:alfvscaling}
\end{figure}

We are now able to fit Equation~\eqref{eq:scaling-alfv} with all our simulations listed in Tables \ref{tab:list} and \ref{tab:list1}. The fit is shown in Figure \ref{fig:alfvscaling} in blue. The value of the fitted parameters are: $K_1= 0.20 \pm 0.004$ and $m=0.41 \pm 0.002$. As in \citet{revi15,revi16}, we find that the open flux law is able to fit well all simulations, independently of the (complex) topology of the surface magnetic field. \cite{revi16} found the fitted parameter values as: $K_1= 0.55 \pm 0.05$, $K_2=0.06 \pm 0.01$ and $m=0.31 \pm 0.02$. Despite the fact that the two sets of coefficients are different, we see in Figure \ref{fig:alfvscaling} that the law of \citet{revi16} remain consistent with our Alfvén wave simulations. In the current study, the exponent $m$ is driven to higher values (and $K_1$ to lower values) because of higher range of the average Alfvén radius distribution. \citet{pant17} have shown that higher coronal base temperature could lead to higher $m$ exponent, and we plot the open flux law they obtained for a coronal base temperature of $2.4$ MK (the base temperature in \citet{revi16} is set to $1.5$ MK). Typical maxmimum temperature in the Alfvén wave model are close to $2.0$ MK and this is fully consistent with our simulation data lying in between the two polytropic open flux laws (orange and green lines) for the angular momentum. This study hence shows that the results obtained with polytropic models and Alfvén waves models are close, as far as angular momentum loss and its scaling with the open flux are concerned.


\begin{figure*}[!ht]
\centering
\begin{tabular}{cc}
\includegraphics*[width=1.05 \linewidth]{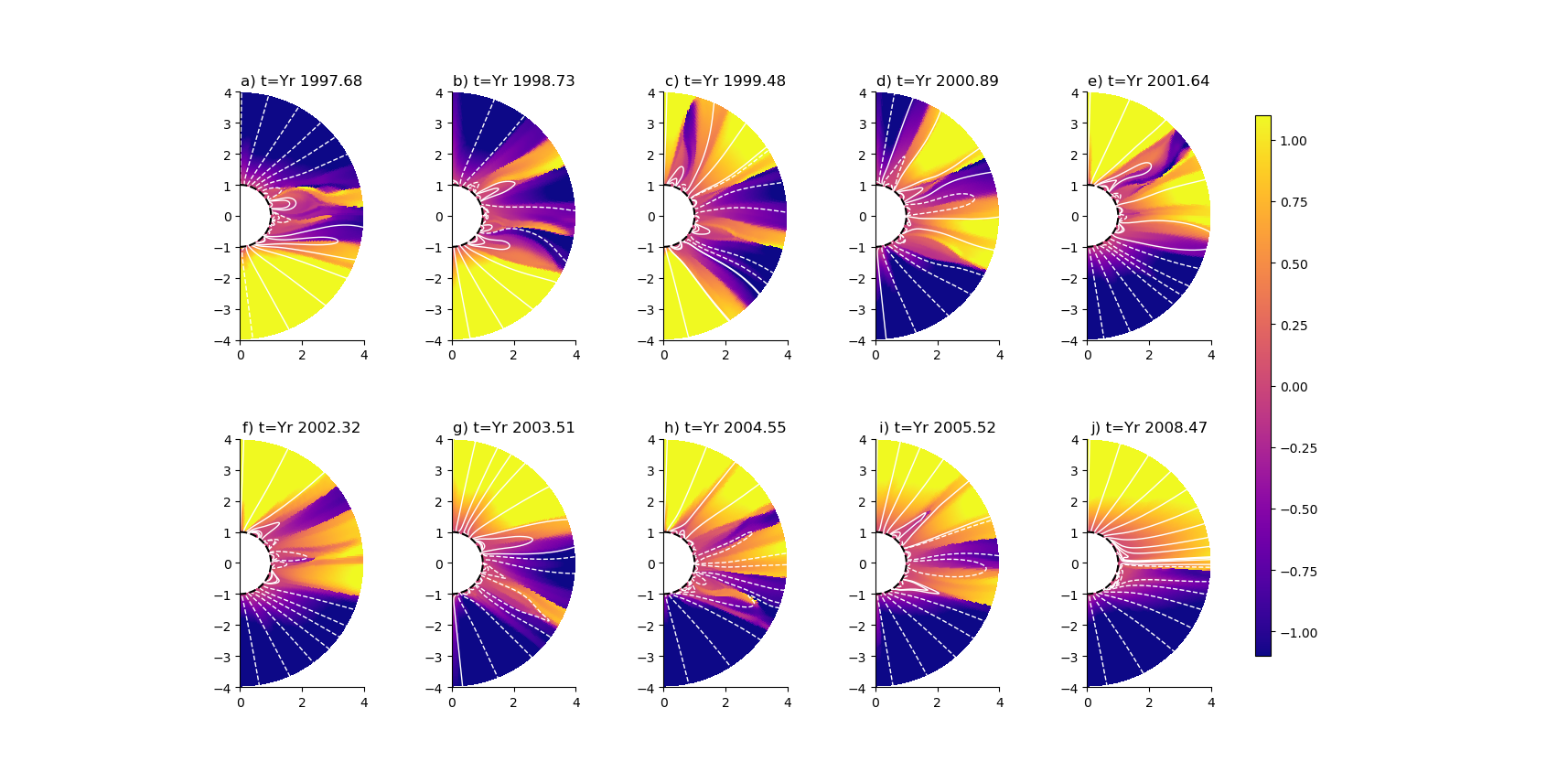}\\
\end{tabular}                
\caption{\footnotesize{Snapshots of the coronal magnetic field evolution with the AW model at different times over cycle 23. Times $t=1997.69, ~1998.73, ~ 1999.48, ~2000.89, ~ 2001.64, ~ 2002.32, ~2003.51, ~2004.55, ~2005.52$ and $2008.47$ are given in fractional years, and start from the first minimum of cycle 23. The color in the background shows the quantity ${\bf{u.B}}/(c_s ||{\bf B}||)$, which is basically the solar wind velocity projected on the magnetic field in units of the Mach number. Outgoing poloidal magnetic field lines are represented by white solid lines and ingoing poloidal magnetic field lines by white dashed lines. Here, we only represent the first 4 solar radii.}}
\label{fig:snapshot}
\end{figure*}

\begin{figure}[!ht]
\centering
\begin{tabular}{cc}
\includegraphics*[width=1.1 \linewidth]{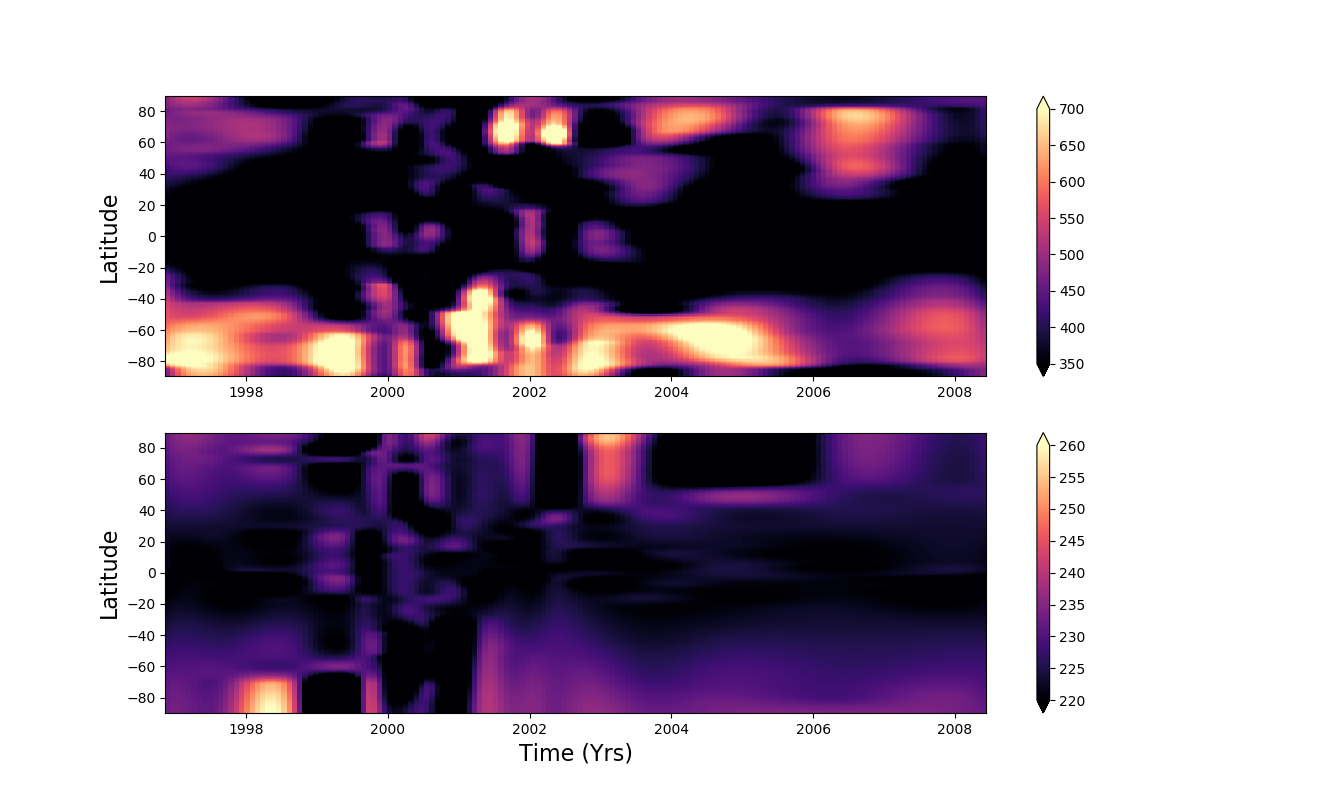}\\
\end{tabular}                
\caption{\footnotesize{Time-latitude diagram of the radial solar wind velocity over the cycle 23 in km/s at $r=20 R_\odot$. Top panel is from the AW model, while the bottom panel is from the polytropic model. Please note the large difference of velocity amplitude breadth in the AW vs polytropic case (350 km/s vs 100 km/s).}}
\label{fig:rad-bfly}
\end{figure}
 
 \begin{table}[!ht]
\caption{Value of the mass loss rate, of the angular momentum (A.M.) loss rate, and of the average Alfvén radius obtained from the polytropic solar wind simulations initialized with the different Carrington rotation maps over cycle 23. }
\label{tab:list2}
\begin{center}
 \begin{tabular}{*5r}
 \hline\hline
 \multicolumn{5}{c}{Polytropic wind model}\\[0.50ex]
  \multicolumn{5}{c}{(Both Axisymmetric and Non-axisymmetric components)}\\[0.50ex]
 \hline
 Carrington Rotation & Fractional Year & Mass Loss & A.M. Loss & Alfvén Radius \\ [0.50ex]
 (CR) & (yr) &($10^{-14}$ $M_\odot/\textrm{yr}$) & ($10^{29}$ erg) & (${\rm R_\odot}$)  \\ 
 
 \hline
	1916 & 1996.86& 2.03&	2.32 & 3.818 \\
	1927 & 1997.68 &1.98 & 2.75& 3.822\\
	1941 & 1998.73 &2.01 & 1.71 & 3.283 \\
	1951 & 1999.48 &1.89 & 0.93 & 2.506\\
	1953 & 1999.63 &1.90 & 1.71&3.384\\
	1961 & 2000.23 &1.65 & 1.61 & 3.512\\
	1965 & 2000.52 &1.91 & 1.38 & 3.031\\
	1970&  2000.89 &1.70 & 1.68 & 3.542\\
	1975 & 2001.27 &2.00& 2.42 & 3.941\\
	1980 & 2001.64 &1.98& 2.39 & 3.926\\
	1985 & 2002.02 &1.97 & 1.90& 3.515\\
	1989 & 2002.32 &1.81 & 1.92 & 3.671\\
	1994 & 2002.69 &1.96 & 1.74 & 3.359\\
	2005 & 2003.51 &2.02 & 1.96 & 3.540\\
	2019 & 2004.55 &1.89 & 1.77 & 3.469\\
	2043 & 2006.35 &2.02 & 1.84 & 3.418\\
	2058 & 2007.47 &2.01& 1.74 & 3.303\\
	2071 & 2008.47 &2.02 & 1.73 & 3.293\\
 \hline 
\end{tabular}
\end{center}
\end{table}

\section{Evolution of the solar corona and its properties during Cycle 23}
\label{sec:loss}

In this section, we perform a quasi-static study of the evolution of the solar corona along cycle 23. We use both polytropic and AW models, and we have selected nineteen WSO synoptic maps at different epochs of the solar cycle 23 for this purpose. We have already discussed the impact of non-axisymmetric components over the total surface energy (see section \ref{sec:initial}) and we will consider them here using Eq.~\eqref{eq:brm}. We show our selected initial magnetic field topologies in Appendix \ref{sec:appendixb}. We fix the other initial coronal parameters, namely density, transverse velocity, and the coronal heating flux for the AW model. We take the value of density as $3 \times 10^8$~cm$^{-3}$, transverse velocity as $30$\,km/s and coronal heating flux $F_h$ as $5 \times 10^4$ erg.cm$^{-2}$ s$^{-1}$. On the other hand, we fix the density of the solar corona to $3 \times 10^8$~cm$^{-3}$ for the polytropic solar wind model. As the initial assumptions and parameters are different for both models, we can not directly compare the result in terms of absolute amplitude. However, we can compare the trends of different physical quantities, namely the mass loss, angular momentum loss, and average Alfvén radius along the cycle 23. Please note that we compare the polytropic and AW model in the previous section when both the models satisfied the same mass loss criteria.

Figure \ref{fig:snapshot} shows the time evolution of the coronal magnetic field at different epochs of the solar cycle 23, zoomed close to the Sun. Different panels of Figure \ref{fig:snapshot} represent different epochs of the solar cycle 23. The color background of Figure \ref{fig:snapshot} represents the quantity  ${\bf{u.B}}/(c_s ||{\bf B}||)$, which is basically the solar wind velocity projected on the magnetic field in units of the Mach number. One will be able to easily identify the change of magnetic field polarity using this quantity. The regions of the magnetic polarity transition correspond to current sheets. Figure \ref{fig:snapshot} clearly shows the changes in the magnetic field topology over the solar cycle 23. At the beginning of solar cycle 23 (panel a), the magnetic field is mostly dipolar. Magnetic field lines are going inward in the northern coronal hole and going outward in the southern coronal hole. As the solar cycle progresses, the surface magnetic field structure becomes more complex, sculpting the corona. We find the occurrence of pseudo streamers at various latitudes as the cycle progresses (see panels b to i of Figure \ref{fig:snapshot}. Coronal holes with magnetic field lines going inward also start to appear in the southern hemisphere as the cycle progresses (panels c and d). Finally, near the end of solar cycle 23, the magnetic field returns again to the almost dipolar but with complete opposite configuration (see panel j of Figure \ref{fig:snapshot}). Magnetic field lines are now fully going inward in the southern coronal hole and going outward in the northern coronal hole.  

Next, we study the variation of wind speed during solar cycle 23. As wind speed is mostly radial at 20 solar radii, we specifically focus on the radial wind speed variations. Top panel of the figure \ref{fig:rad-bfly} shows the radial wind speed evolution at 20 solar radii during the solar cycle 23. At the beginning phase of the solar cycle, fast wind originates from the high latitudes, while slow wind originates at the lower latitudes. As the solar cycle progresses, the slow wind starts to appear at higher latitudes, and fast wind structure is also observed at lower latitudes. At the end phase of the solar cycle, again fast wind appears at high latitude and slow wind at low latitudes. The time-latitude diagram of the radial wind speed (top panel of the figure \ref{fig:rad-bfly}) also indicates that wind speed is often asymmetric. This is due to the presence of strong hemispheric asymmetries in the surface magnetic field of individual synoptic maps during solar cycle 23. The southern hemisphere is more active compare to the northern hemisphere during solar cycle 23. We also notice sometimes there is a significant drop in the wind velocity at the poles in the two hemispheres. \cite{khab17} interpreted this kind of drop (also observed in the Ulysses data) as the presence of high-latitude current sheets. The bottom panel of the figure \ref{fig:rad-bfly} shows the same obtained from the polytropic wind model. Although we find the behaviour of radial wind speed variation in the polytropic scenario is almost similar to the AW one, the distribution breadth of the wind speed is in a much narrower range compared to the AW one. 

\begin{figure}[!ht]
\centering
\begin{tabular}{cc}
\includegraphics*[width=\linewidth]{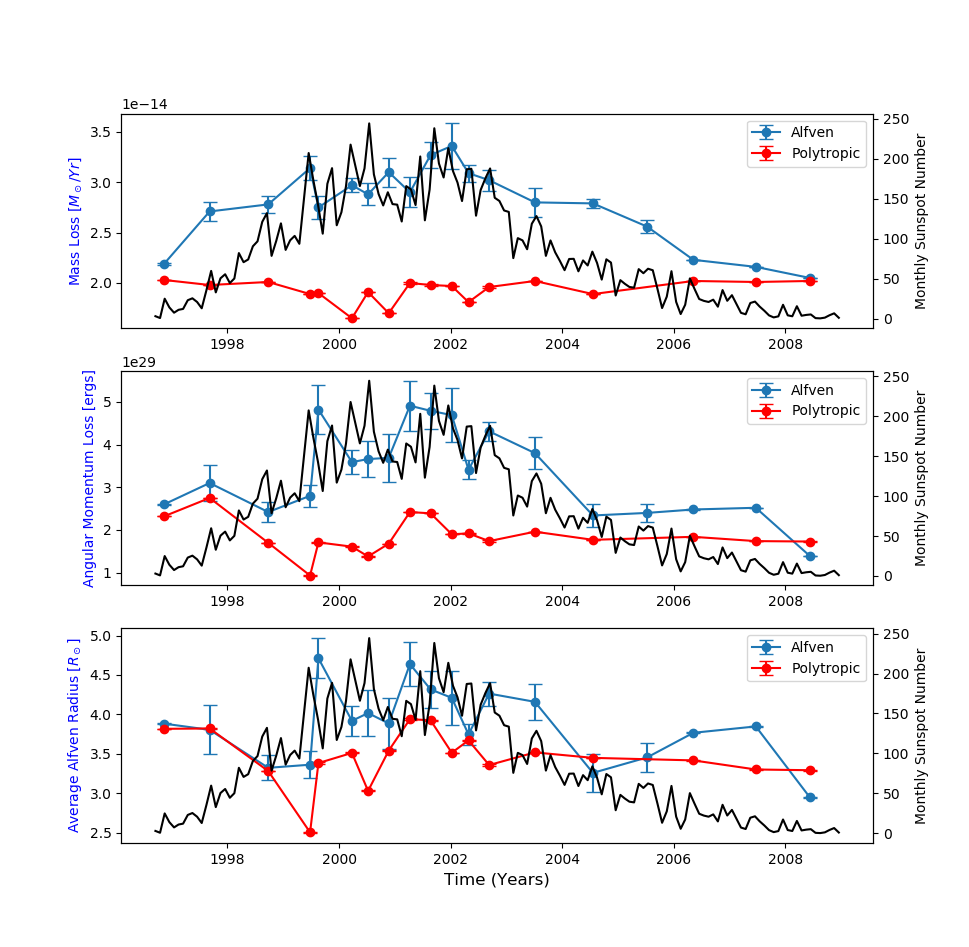}\\
\end{tabular}                
\caption{\footnotesize{Time evolution of the mass loss, angular momentum loss and the average Alfvén radius during cycle 23 obtained from the polytropic and Alfvén wave driven wind simulations. Blue lines correspond to the polytropic wind simulations, while red lines corresponds to the Alfvén wave driven wind simulation. Black lines represents the monthly sunspot number during the cycle 23. Here, we have included the contribution of non-axisymmetric components in both axisymmetric polytropic and Alfvén wave driven wind simulations with the observed magnetic maps following the quadrature addition method (equation \ref{eq:brm}). }}
\label{fig:non-axi-time}
\end{figure}

Finally, we focus on the time evolution of different integrated quantities, namely, mass loss (Eq. \ref{eq:ml}), angular momentum loss (Eq. \ref{eq:jl}), and the average Alfvén radius (Eq. \ref{eq:ra}) over the solar cycle 23. Their evolution is displayed in Fig. \ref{fig:non-axi-time} for polytropic (red circles) and AW (blue circles) models. We have also overplotted the variation of monthly sunspot numbers for a better understanding of our results.

A large number of studies have been performed to estimate the solar mass loss and angular momentum loss rates along the solar cycle \citep{usma20, rile01, pint11, revi17, perr18}. These earlier studies have reported a wide range of values for solar mass loss and angular momentum loss rates. This is probably due to the fact that all of these studies have different input physics, like the inclusion of specific coronal heating functions or the polytropic assumption. However, we note that in all these models it is possible to reproduce the observed solar mass loss by adjusting the input parameters. In this study, we have estimated the mass loss and angular momentum loss rates from both the AW and polytropic solar wind models. Table \ref{tab:list1} and \ref{tab:list2} lists the value of different physical quantities obtained from both models along cycle 23. Our calculated mass loss from the AW model varies from $ 2.1 \times 10^{-14} ~M_\odot/\textrm{yr}$ at the solar minimum to $ 3.5 \times 10^{-14} ~M_\odot/\textrm{yr}$ at the solar maximum (see top panel of Figure \ref{fig:non-axi-time}; blue curve). Average mass loss over solar cycle 23 using this Alfvén wave driven wind simulation is $ 2.8 \times 10^{-14} ~ M_\odot/\textrm{yr}$. Our calculated mass loss is minimum at the minimum of the solar cycle and maximum at the maximum of the solar cycle. \cite{mcco08} estimated the mass loss from the Ulysses spacecraft data and found a maximum mass loss value of $3.1 \times 10^{-14} ~ M_\odot/\textrm{yr}$ during 1992-93 and a minimum value of $2.3 \times 10^{-14} ~ M_\odot/\textrm{yr}$ during 1991. Our minimum value of the mass loss is approximately the same as the observed mass loss value; while the maximum value is higher than the observed mass loss value. We found almost $75 ~\%$ variation in the mass loss between the solar cycle minimum and maximum which is quite larger than earlier studies. Finally, we note that the positive correlation between the mass loss and the solar cycle in the AW model stems from the fact that the Poynting flux directly depends on the surface magnetic energy of the magnetic maps.

On the other hand, we find a completely different behaviour in the polytropic scenario. The top panel of Figure \ref{fig:non-axi-time} indicates that the mass loss weakly varies along the cycle in the polytropic scenario (red curve), with a slight decrease for the first three years and a slight increase during the descending phase of the cycle. The mass loss varies from $1.65 \times 10^{-14} ~ M_\odot/\textrm{yr}$ to $ 2.1 \times 10^{-14} ~ M_\odot/\textrm{yr}$ (variation is at most 30 \%) in the polytropic scenario. In recent times, different studies have included the effect of magnetic variability in their wind simulations. \cite{pint11} and \cite{perr18} used a kinematic solar dynamo model to drive their axisymmetric polytropic solar wind simulations. Both of these studies found a smaller variation compared to our AW model between the mass loss rates obtained from the solar maximum and minimum respectively. We also find a lower mass loss variation in the polytropic scenario compared to the AW one. In our study the polytropic model shows an anti-correlation between mass loss and solar cycle because a larger magnetic energy generally means that more close loops can be formed in the polytropic approximation, and thus the mass loss can be expected to be smaller at cycle maximum. \citet{pint11} nevertheless found a mass loss is correlated with the solar cycle, due to the surface energy-magnetic cycle phase correlations in the Babcock-Leighton dynamo models they used. Likewise, \citet{perr18} found yet another behaviour with a phase quadrature correlation, that comes from the phase quadrature between cycle phase and surface magnetic field energy embedded in their $\alpha$-$\Omega$ dynamo model. When using observed magnetograms, we thus show here that the polytropic model is not good enough and tends to predict a probably wrong anti-correlation between cycle phase and mass loss.

The middle panel of Figure \ref{fig:non-axi-time} shows the time evolution of the angular momentum loss during the solar cycle 23 from both AW and polytropic solar wind models. In the AW model, angular momentum loss is correlated with the solar cycle (see middle panel of Figure \ref{fig:non-axi-time}; blue curve). Angular momentum loss varies from $10^{29}$ erg at the solar minimum to the $5.2 \times 10^{29}$ erg at the solar maximum. The variations in the angular momentum loss is also higher compared to earlier studies by \citep{pint11} and \cite{perr18} (see middle panel of Figure \ref{fig:non-axi-time}). The average angular momentum loss obtained from our AW model during the solar cycle 23 is around $4 \times 10^{29}$ erg. On the other hand, we found a anti-correlation of the angular momentum loss with the solar cycle in the polytropic wind model scenario (see middle panel of Figure \ref{fig:non-axi-time}; red curve). It means the angular momentum loss rate is minimum when the star is more multipolar. Earlier studies based on polytropic wind model also found the same (see \cite{pint11}, \cite{perr18}). We also notice larger angular momentum loss variation in the AW model compared to the polytropic one.

The last panel of Figure \ref{fig:non-axi-time} shows the time evolution of the Alfvén radius during the solar cycle 23 from both models. In the AW model, we find a maximum average Alfvén radius of around 4.8 solar radii at the cycle maximum, while the minimum average Alfvén radius is around 3 solar radii at the cycle minimum. We note that our average Alfvén radius is smaller compared to what one can expect from Helios mission. \cite{pizz83} and \cite{mars84} estimated the average Alfvén radius from the angular momentum obtained from the Helios mission. They found the value of the average Alfvén radius as 12-13.6 solar radii at the cycle minimum and 13.6-16.6 solar radii at the cycle maximum. However, we find different time evolution behaviour of the Alfvén radius in the polytropic scenario. The Alfvén radius decreases in the first few years of the solar cycle (see bottom panel of Figure \ref{fig:non-axi-time}; red curve) . \cite{pint11} and \cite{perr18} found that the Alfvén radius is anti-correlated with the solar cycle; that means the Alfvén radius is minimum at the cycle maximum and minimum at the cycle maximum. However, \cite{revi17} used a 3D wind model based on polytropic assumption and found that the Alfvén radius increased from $5 {\rm R_\odot}$ at the cycle minimum to the $7 {\rm R_\odot}$ at the cycle maximum. We note that the average Alfvén radius is closely related to the ratio of $\dot{J}/\dot{M}$. As a matter of fact both $\dot{J}$ and $\dot{M}$ can vary differently along the solar cycle depending on the details of the wind model. It can thus be expected that the average Alfvén radius may show positive and negative correlations with the magnetic cycle depending on the modelling details.

Finally, we note that our AW model predicts significantly different trends compared to wind models based on the polytropic assumption. In principle, this difference could originate either from the physical model itself (polytropic and AW assumption), or from our choice of summing quadratically the non-axisymmetric contributions in our 2.5D approach. To test this, we also perform 2.5D simulations considering only the axisymmetric components of the surface magnetic field. One can see Appendix \ref{sec:appendixa} for a discussion regarding the wind simulations where we consider only the axisymmetric components of the surface magnetic field. Specifically, we first note that the AW model predicts almost no variation of mass loss along the magnetic cycle when considering axisymmetric modes only (see Figure \ref{fig:axi-time}). This behaviour then becomes more similar to the one found with the polytropic model (Figure \ref{fig:non-axi-time}), which shows very similar trends whether or not non-axisymmetric modes are taken into account. Interestingly, in this case the angular momentum loss also becomes almost flat along the solar cycle, and thus the Alfvén radius trend is actually very similar with and without non-axisymmetric modes in the AW model. We therefore conclude that the difference of trends between the AW and polytropic model comes from a fundamental difference in how both models depend on the magnetic structure at the surface. Indeed, the surface magnetic field topology contributes directly to the input energy source in the AW model through the Poynting flux, which is not the case for the polytropic model where the wind is mostly driven by a fixed pressure gradient in our study. That is why the Alfvén wave driven wind model is much more sensitive to the complex surface magnetic field topology compared to the polytropic one.

\section{Summary and Conclusions}
\label{sec:summary}

We have performed this study to understand the behaviour and properties of the solar wind obtained from an Alfvén wave driven solar wind model. We have compared the Alfvén wave driven wind model results with a polytropic approach. We have found significant differences in the behaviour and properties of the solar wind obtained from these two models. We find that Alfvén wave driven model is able to reproduce both the observed speed bi-modality and the observed speed distribution breadth of the solar wind, while polytropic model is only able to produce a weakly pronounced bi-modality. Our study reconfirms the necessity of an additional acceleration mechanism (like Alfvén waves dissipation) on top of the thermal one in a wind model to explain the observed speed bi-modality and distribution breadth, as predicted by earlier studies. Alfvén waves act here as additional acceleration and heating mechanisms in our model which helps reproducing the observed bimodality. 

Although a polytropic wind model can not explain the observed bimodality, both polytropic and Alfvén wave driven wind models show similar kinds of wind speed distribution (slow wind near streamers and the fast wind away from streamers). Away from streamers, open magnetic field lines allow the wind to achieve comparatively faster speed. Some earlier studies also suggested that the open magnetic field in the cooler regime (coronal hole) of the corona reconnects with the closed magnetic loops near the base, imparting extra energy into the overlying corona. This extra energy can act as an additional acceleration mechanism and increases the wind speed in the cooler (coronal hole) regime of the solar corona \citep{fisk03, gloe03, schw14}. 

One interesting observational feature is the anti-correlation between the final solar wind velocity and the coronal source temperature \citep{geis95}. However, thermally driven, polytropic winds show a positive positive correlation between the final wind velocity and the coronal source temperature which is against observations (see Figure \ref{fig:veltemp1}). We have investigated this issue for our Alfvén wave driven wind model. We find a clear anti-correlation between the final wind velocity and the coronal source temperature. This result is probably due to the consequence of strong temperature dependence of the thermal conduction term or due to the different radial distribution of heating rate as suggested before \citep{schw03, cran05, cran10, vand10}. However, we get a positive correlation between the final wind velocity and the temperature from the AW model if we consider the temperature away from the bottom coronal boundary. Shifting from negative to positive correlation in the AW model occur at different heights of the solar corona depending on the assumed initial coronal parameters and the surface magnetic field.

We find that the global properties of the modeled wind, like the mass loss rate, the angular momentum loss rate, and the average Alfvén radius, are sensitive to the different coronal parameters. Specifically, the surface magnetic field topology plays an important role in the variation of the global properties through the Poynting flux. We also proposed a single scaling law to compute the mass loss (Eq. \ref{eq:scaling-massloss}) and the average Alfvén radius (Eq. \ref{eq:scaling-alfv}) in Alfvén wave driven wind models. We find that the mass loss scaling law is very different from the one derived with polytropic approach, but the average Alfvén radius scaling-law \citep{revi15a,revi16} seems actually relatively independent of the wind model.  Finally, we note that that mass loss is positively correlated with the initial surface magnetic field strength in the Alfvén wave driven wind model.

The evolution of different magnetic modes of the surface magnetic field over an eleven-years solar cycle induces a different type of complexity in the solar corona at different epochs of the solar cycle. The surface magnetic field structure at the minimum of the solar activity is mostly dipolar. We also find that axisymmetric components of the surface magnetic fields are the major contributor to the total surface energy at the activity minimum. As the solar wind carries the topological information of the surface magnetic field, one can expect less complexity in the coronal structure at the activity minimum. We also find a very organized behaviour of the solar wind at the cycle minimum --- slow wind at the equator and the fast wind at the high latitudes. However, the surface magnetic field is much more complex and multipolar at the cycle maximum. On the contrary to activity minimum, non-axisymmetric components are the major contributor to the total surface energy at the cycle maximum and we find much more complex wind distribution at the cycle maximum. This is consistent with Ulysses observations that revealed similar structures in the solar wind \citep{mcco08}.

As the solar corona evolves over the eleven-years solar cycle, different global wind properties are also expected to evolve. We find that the mass loss rate and the angular momentum loss rate are strongly correlated with the solar cycle, while this is not the case with the polytropic wind model. Previous studies based on the polytropic wind model found a weak anti-correlation of the angular momentum loss rate with the solar cycle \citep{pint11, perr18}. We also find a large difference between the physical quantities calculated at the cycle minimum and maximum from the Alfvén wave driven wind model when we considered the non-axisymmetric components of the surface magnetic field. However, this difference is much less when we consider only axisymmetric components of the surface magnetic field. These differences between the polytropic and the Alfvén wave driven wind models are basically due to the difference in the fundamental assumption of these models. In the polytropic wind model, there is no direct contribution of the magnetic field on the input energy source. Input energies mostly come from internal energy through the pressure gradient. However, in the Alfvén wave driven wind model, the magnetic field enters directly into the input energy sources through the Poynting flux. That's why different physical quantities are much more sensitive to the assumed surface magnetic fields in the Alfvén wave driven solar wind model.

Finally, we believe that Alfvén wave driven wind model is a step toward understanding the global properties of the solar corona in a better way. Please note that we performed this study in 2.5D, assuming axisymmetry, yet accounting for the contribution of non-axisymmetric energy. It nevertheless lacks the information of the full three-dimensional topology of the solar surface magnetic field. Consideration of the full three-dimensional topology would bring additional information and help us to better understand the global properties of the solar wind. We leave this aspect for our future studies.

\acknowledgements
We thank Rui Pinto and Susanna Parenti for useful discussion and suggestions. We thank the Université Paris-Saclay (IRS SPACEOBS grant), ERC Synergy grant 810218 (Whole Sun project), INSU/PNST and CNES Solar Orbiter for supporting this research. We acknowledge access to supercomputers through GENCI (grants 40410133, 60410133, and 80810133 ).

\begin{table}[!ht]
  \caption{Mass loss for the solar cycle 23 maximum (CR 1975) and minimum (CR 1919)}
  \label{tab:massloss}
  \begin{center}\begin{tabular}{ccccc}
    \hline\hline
     & \multicolumn{2}{c}{CR 1975} & \multicolumn{2}{c}{CR 1919} \\ \hline
Mass Loss ($M_\odot/\textrm{yr}$)    & Only Axisymmetric & Considering contribution & Only Axisymmetric & Considering Contribution  \\
  & Components& of Non-Axisymmetric & Components& of Non-axisymmetric \\
 & & Components& & Components\\
  \hline
Polytropic Wind Simulation & $2.95 \times 10^{-14}$ & $2.75 \times 10^{-14}$&$2.67 \times 10^{-14}$&$2.67 \times 10^{-14}$\\
Alfvén wave driven wind simulation &$2.14 \times 10^{-14}$&$2.68 \times 10^{-14}$&$2.38 \times 10^{-14}$&$2.37 \times 10^{-14}$\\
\hline \\
\end{tabular}
  \end{center}
\end{table} 

\appendix

\section{Influence of Non-Axisymmetric Surface Magnetic field components on the Wind solutions}
\label{sec:appendixa}

\begin{figure}[!ht]
\centering
\begin{tabular}{cc}
\includegraphics*[width=\linewidth]{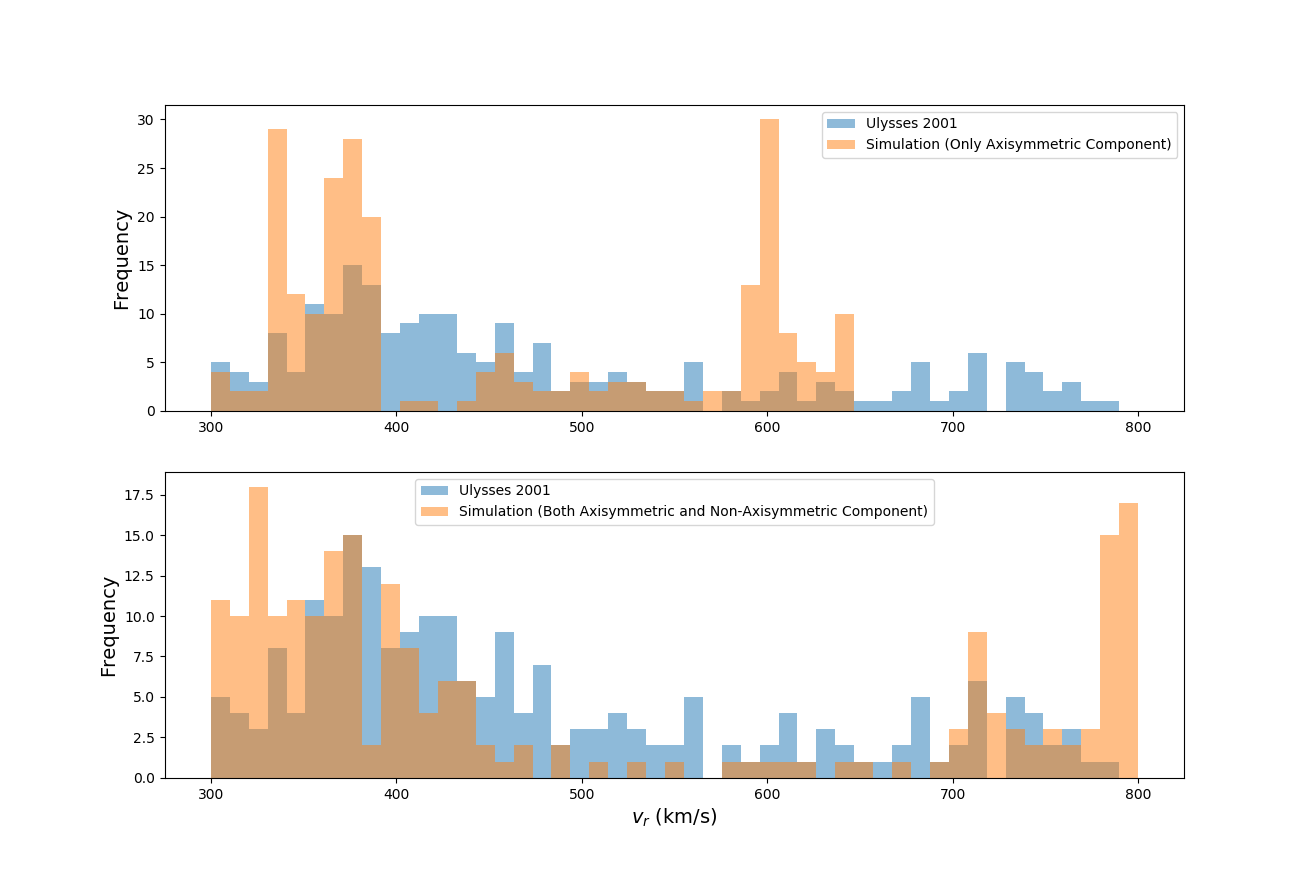}\\
\end{tabular}                
\caption{\footnotesize{Histograms of the radial wind velocity obtained from the Ulysses data (blue color) and our Alfvén wave driven wind simulations (orange color). For the figure in the top panel, we perform the axisymmetric wind simulations considering only the axisymmetric components of initial magnetic map. While for the figure in the bottom panel, we have performed the simulations considering the contribution of non-axisymmetric components following equation \ref{eq:brm}.}}
\label{fig:hist1}
\end{figure}

\begin{figure}[!ht]
\centering
\begin{tabular}{cc}
\includegraphics*[width=1.05 \linewidth]{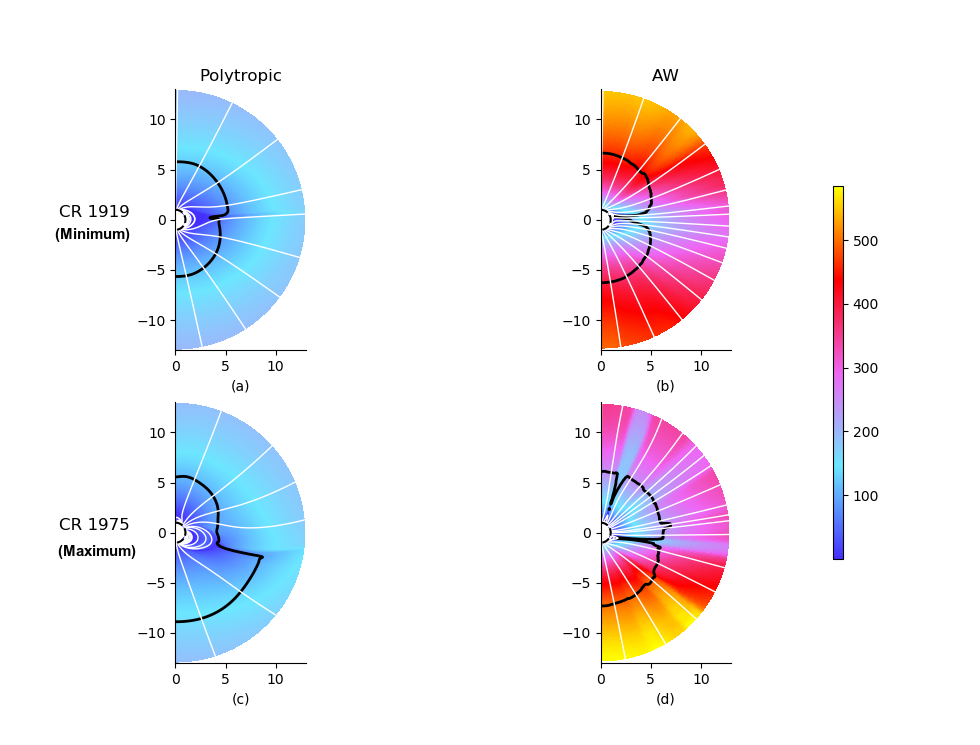}
\end{tabular}                
\caption{\footnotesize{Same as in Figure \ref{fig:alfv1}. However, here we perform the wind simulations considering only the axisymmetric components.}}
\label{fig:alfv2}
\end{figure}

\begin{figure}[!ht]
\centering
\begin{tabular}{cc}
\includegraphics*[width= 1.05 \linewidth]{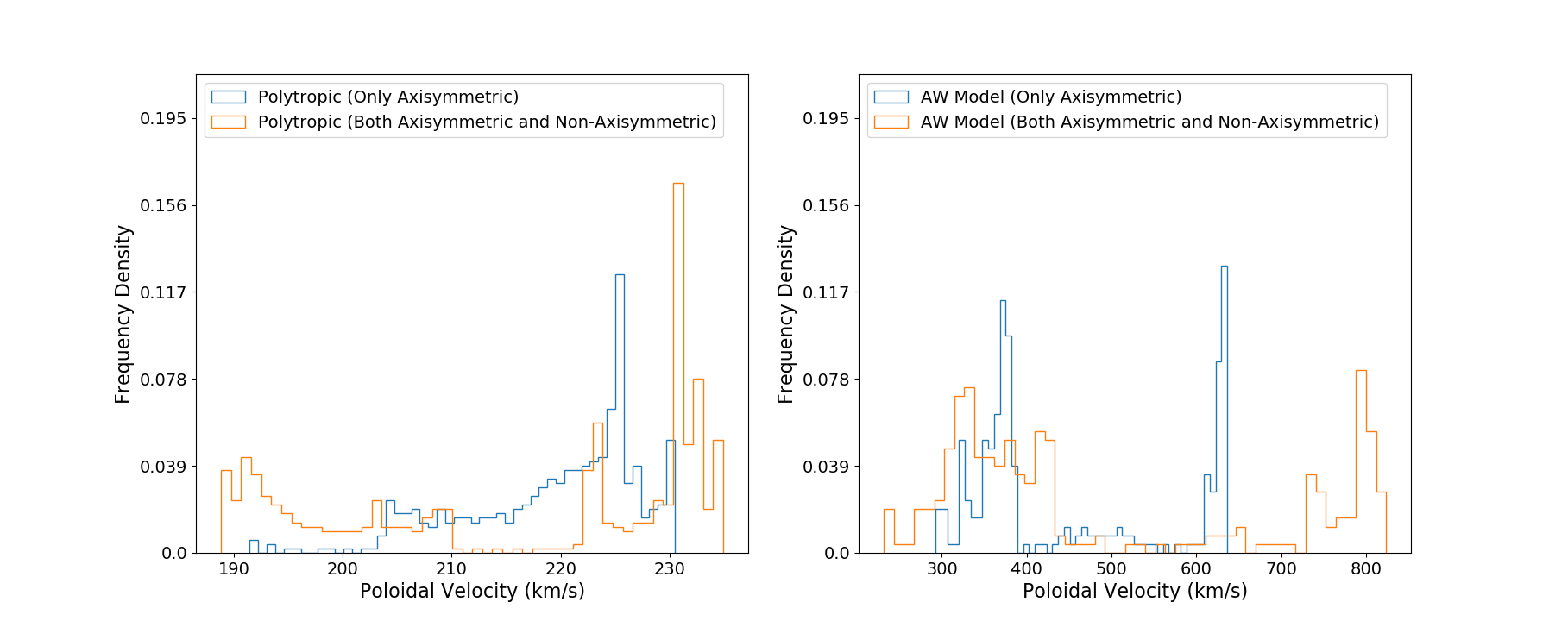}
\end{tabular}                
\caption{\footnotesize{Histograms of the radial wind velocity obtained from the both polytropic and Alfvén wave driven wind simulations. Left panel figure is obtained from the polytropic wind simulations; while right panel figure is from Alfvén wave driven wind simulations. In both category, we consider two scenarios. In one scenario, we initialize the wind model considering only axisymmetric components of the decomposed surface magnetic field obtained from the magnetic map CR 1975. In another scenario, we perform the wind simulation considering the contribution of non-axisymmetric components following equation \ref{eq:brm}. In both category (polytropic and Alfvén wave), we find higher poloidal velocity when we consider the contribution of non-axisymmetric components.}}
\label{fig:hist3}
\end{figure}

\begin{figure}[!ht]
\centering
\begin{tabular}{cc}
\includegraphics*[width=\linewidth]{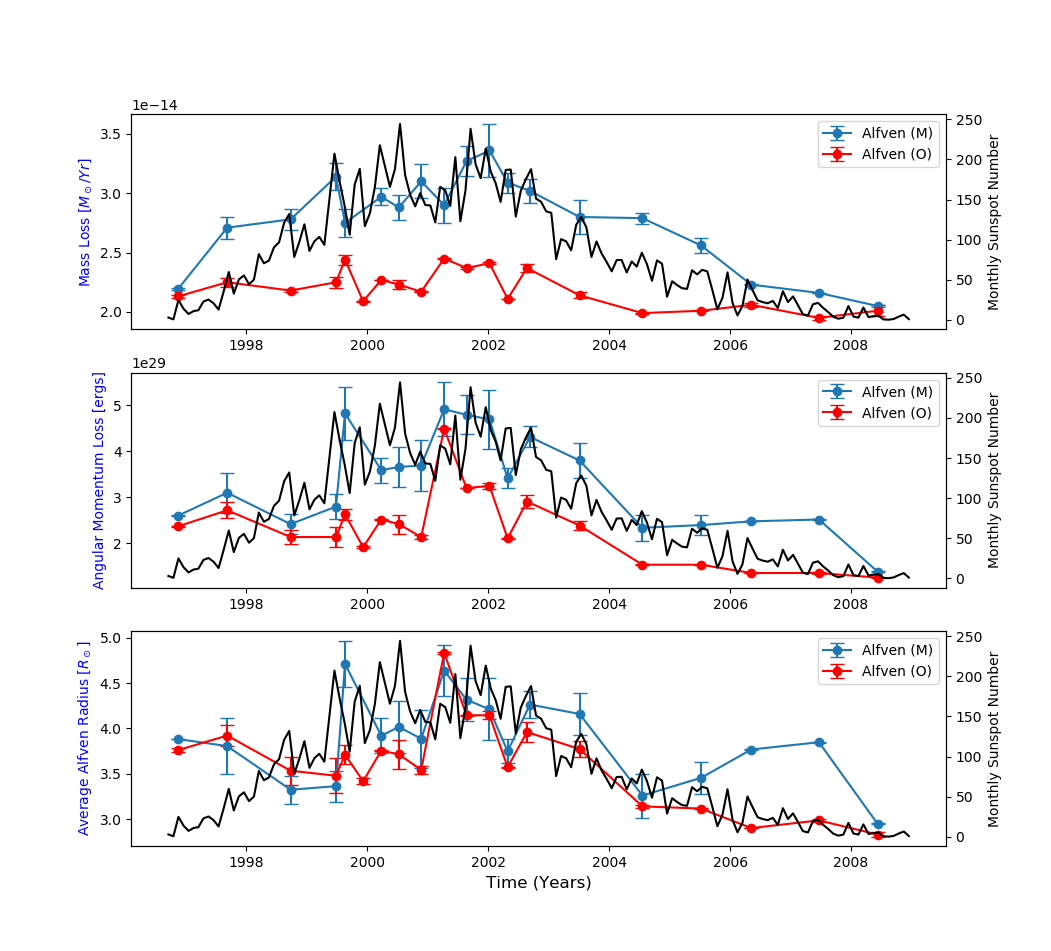}\\
\end{tabular}                
\caption{\footnotesize{Time evolution of the mass loss, angular momentum loss, and the average Alfvén radius during cycle 23 obtained from the two Alfvén wave driven wind simulation scenarios. Red line [Alfvén (O)] corresponds to the scenario where we initialize the wind model considering only axisymmetric components of the decomposed surface magnetic fields. Blue line [Alfvén (M)] corresponds to the scenario where we perform the wind simulations considering the contribution of non-axisymmetric component following the quadrature addition method (Equation \ref{eq:brm}). We get significantly higher mass loss and angular momentum loss at the maximum of the solar cycle when we consider the contribution of non-axisymmetric component.}}
\label{fig:axi-time}
\end{figure}

 We followed two approaches to model the solar wind along the solar cycle with the AW and polytropic models. In one scenario, we perform axisymmetric Alfvén wave driven wind simulation with decomposed surface magnetic field data obtained from the CR 1975 where we have not included the contribution of non-axisymmetric components. In the other scenario, we perform the same set of simulations but considered the contribution of non-axisymmetric components following Equation~\eqref{eq:brm}. 
 
 First, let us compare the radial velocity variations at 0.1 AU obtained with the AW models with the observed Ulysses radial velocity variation. Figure \ref{fig:hist1} shows the comparison which suggests that we can explain the observed Ulysses radial velocity variation in a better way when we consider the contribution of both axisymmetric and non-axisymmetric components of the surface magnetic fields to initialize the wind simulations (lower panel).

We aim next to characterize the impact of non-axisymmetric surface magnetic field components on the wind solutions obtained from the both Alfvén wave driven wind model and polytropic solar wind model. Figure \ref{fig:alfv2} (a) and (b) shows the poloidal velocity in the meridional plane obtained from the polytropic and Alfvén wave driven solar wind model when initialized with the only decomposed axisymmetric components of the surface magnetic field for the map of CR 1975. Figure \ref{fig:alfv2} (c) and (d) represents the same when the models are initialized with the decomposed axisymmetric field components for the map of CR 1919. The figure is the same as Fig. \ref{fig:alfv1}, albeit now we consider only the axisymmetric modes of the syoptic maps. We obtain mass loss of around $2.14 \times 10^{-14} \ M_\odot/\textrm{yr}$ and $2.38 \times 10^{-14} \ M_\odot/\textrm{yr}$ from the Alfvén wave driven solar wind model when initialized with the CR 1975 map and CR 1919 map respectively. On the other hand, we get the mass loss of around $2.95 \times 10^{-14} \ M_\odot/\textrm{yr}$ and $2.67 \times 10^{-14} \ M_\odot/\textrm{yr}$ from the polytropic solar wind model when initialized with the CR 1975 map and CR 1919 map respectively. One can see Table \ref{tab:massloss} for the detailed summary of mass loss in all scenarios. 

Interestingly, we notice that the value of mass loss increases at the solar maximum (CR 1975 case) and remains the same at the cycle minimum (CR 1919 case) in the polytropic scenario when we only considered the contribution of axisymmetric components. However, we find a decrease in the mass loss at the solar maximum (CR 1975 case) and almost the same mass loss at the cycle minimum (CR 1919 case) in the Alfvén wave driven model scenario. We have already described that non-axisymmetric components is not very significant at the minimum of the solar cycle; that's why we get the same mass loss at the cycle minimum in both cases. However, the trend of the mass loss change at the solar maximum is opposite in the polytropic and Alfvén wave driven model scenario when we compare the models with and without the contribution of non-axisymmetric components (see Table \ref{tab:massloss}). This is the case even though we find the higher poloidal velocity for both the polytropic and Alfvén wave driven wind model scenario when we consider the contribution of non-axisymmetric components (see Figure \ref{fig:hist3}). 

We finally characterize the impact of non-axisymmetric components on the evolution of physical quantities obtained from our Alfvén wave driven wind simulations along the solar cycle. Figure \ref{fig:axi-time} shows the time evolution of mass loss, angular momentum loss, and the average Alfvén radius obtained from the Alfvén wave driven wind models. In one scenario [Alfvén(O)], we perform the simulations where we have not considered the contribution of the non-axisymmetric component. In the other scenario [Alfvén(M)], we perform the wind simulations considering the contributions of non-axisymmetric components following the quadrature addition method (equation \ref{eq:brm}). We find that mass loss is correlated with the solar cycle in both cases. However, we found a much smaller variation in the mass loss (only 25 \%) between the cycle maximum and minimum in the Alfvén (O) scenario (red curve) compared to Alfvén (M) scenario (blue curve). There are almost 75 \% variations in the mass loss between the cycle maximum and minimum in the Alfvén (M) scenario (blue curve). We note that mass loss is almost the same at the cycle minimum in both scenarios; while it increases more at cycle maximum when we consider the contribution of non-axisymmetric components. This is because non-axisymmetric components play a significant role in the cycle maximum compared to the cycle minimum. Similarly, we also notice that variation of the angular momentum loss between the cycle maximum and minimum is much less in the model where we have not considered the contribution of non-axisymmetric components (Alfvén (O) scenario). The middle panel of Figure \ref{fig:axi-time} also indicates a slightly decreasing trend of the angular momentum loss in the Alfvén (O) scenario at the beginning of the solar cycle. The last panel of Figure  \ref{fig:axi-time} indicates that Alfvén radius is maximum at the cycle maximum and minimum at the cycle minimum in both scenarios. On one side, the inclusion of non-axisymmetric components has a significant impact on the time evolution of the mass loss and angular momentum loss. On the other side, the average Alfvén radius is only mildly affected. We have previously shown that one will underestimate the total surface energy significantly at the solar maximum if they do not consider the contribution of non-axisymmetric components. Thus, it is necessary to include the contribution of non-axisymmetric components.

\begin{figure*}[!ht]
\centering
\begin{tabular}{cc}
\includegraphics*[width=\linewidth]{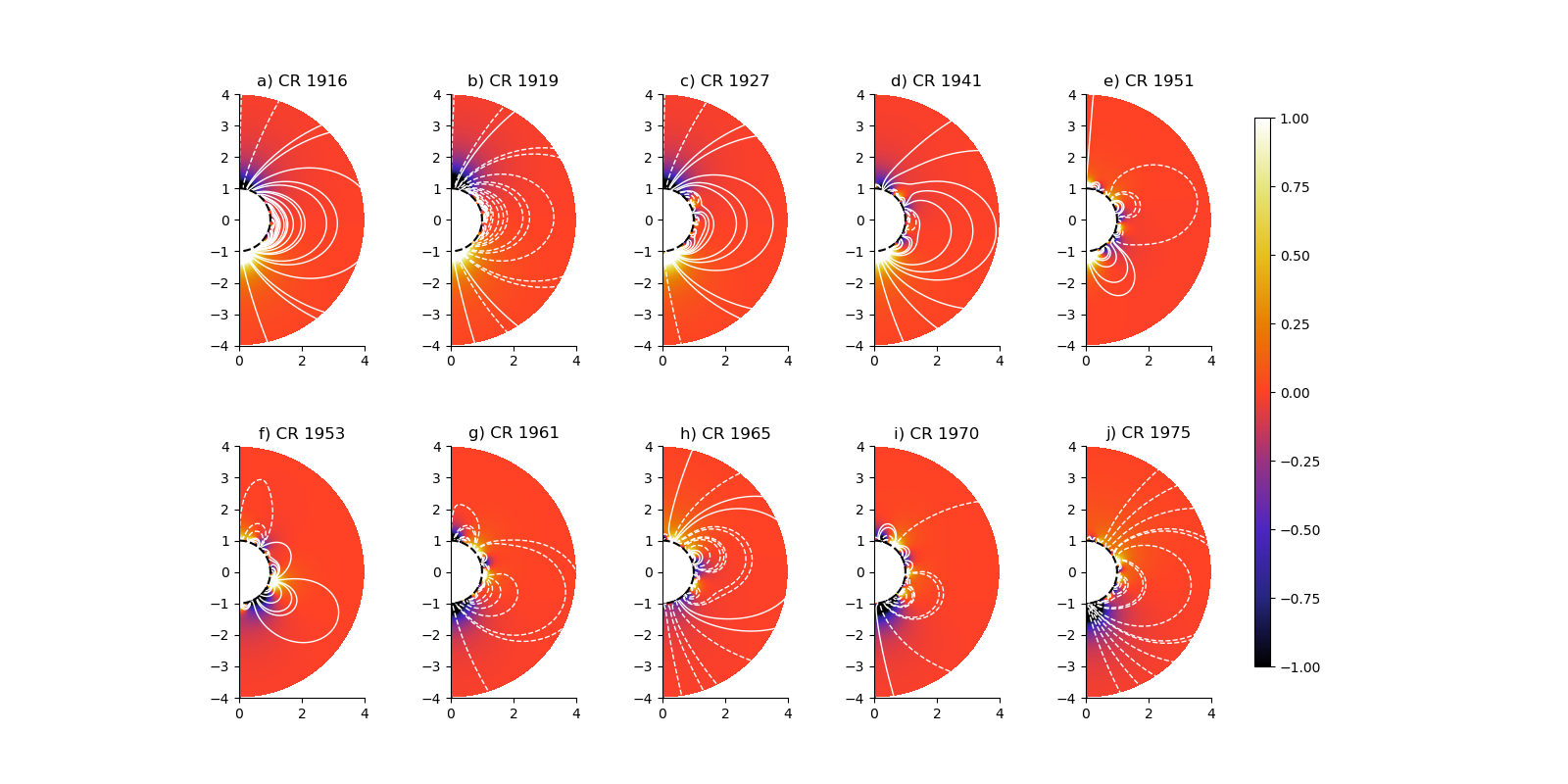}\\
\includegraphics*[width=\linewidth]{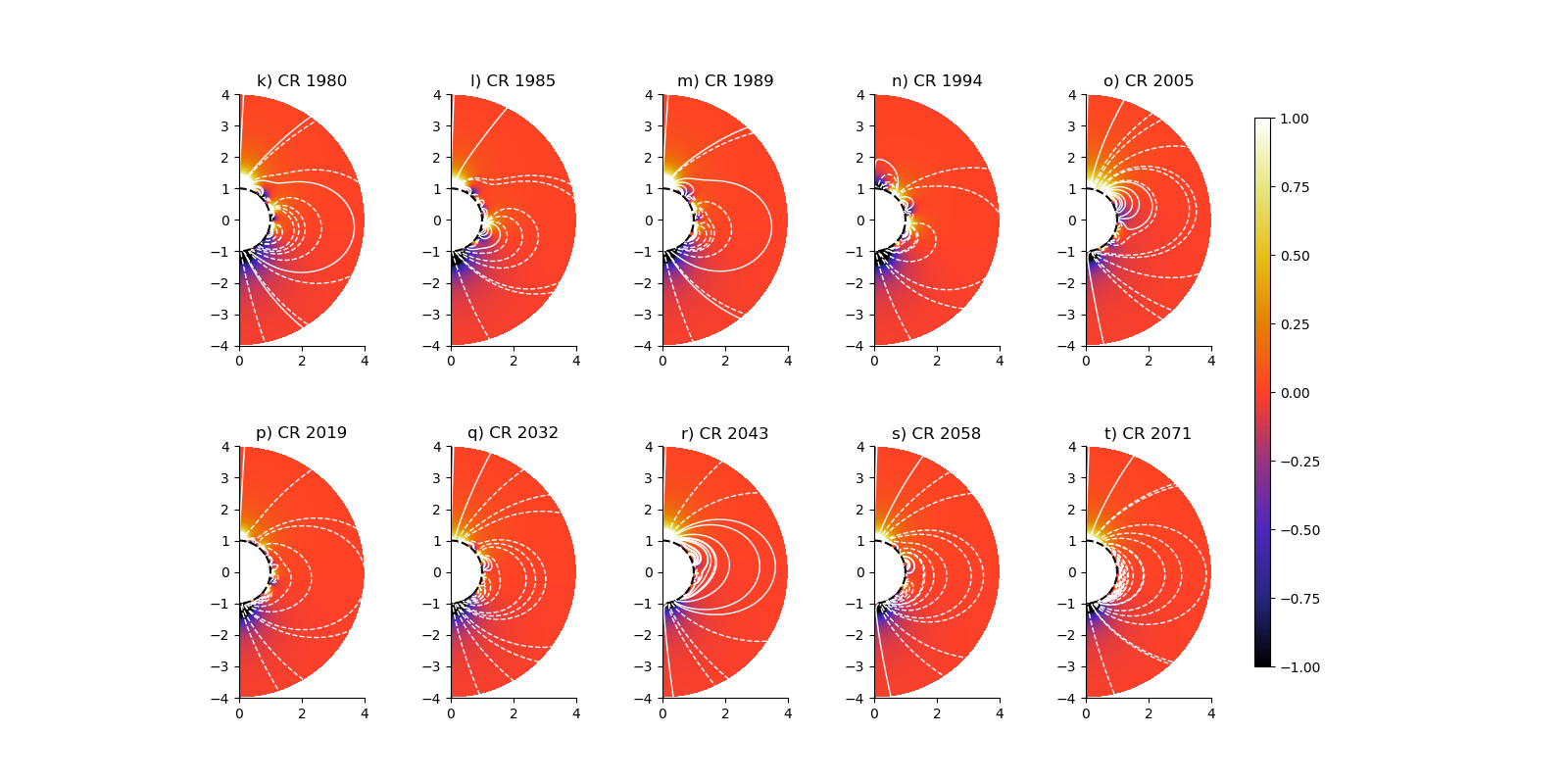}\\
\end{tabular}                
\caption{\footnotesize{Magnetic field lines of our initial topologies obtained from our selected Carrington Rotation (CR) maps during the solar cycle 23. The color in the background shows the radial magnetic field ($B_r$) in G. Outgoing poloidal magnetic field lines are represented by white solid line and ingoing poloidal magnetic field lines by white dashed line. We notice that initial magnetic field topologies are dipolar at the minimum of the solar cycle; while we notice complex magnetic topology at the maximum of the solar cycle. Here, we only represent the first 4 solar radius.}}
\label{fig:snapshot-initial}
\end{figure*}

\section{Initial magnetic field topologies for our selected magnetic maps}
\label{sec:appendixb}
In this appendix, we show the initial magnetic field topologies from our selected Carrington rotation (CR) maps during the solar cycle 23. Figure \ref{fig:snapshot-initial} shows how the initial magnetic field topology (computed with potential field extrapolation with a source surface) varies during cycle 23. At the maximum of the solar cycle (panel i, j, k), we find a complex magnetic field topology. On the other hand, the magnetic field structure is mostly dipolar at the minimum of the solar cycle. Please note that we consider the contribution of both axisymmetric and non-axisymmetric components following Equation \eqref{eq:brm} to get these initial magnetic field topologies. 
\bibliographystyle{yahapj}
\bibliography{reference}

\end{document}